\documentclass{aa}  
\usepackage{graphicx}
\usepackage{txfonts}
\usepackage{color}
\usepackage{lscape}
\usepackage{hyperref}
\begin{document}

   \title{Stellar Population Astrophysics (SPA) with TNG\thanks{Based on observations made with the Italian Telescopio Nazionale Galileo (TNG) operated on the island of La Palma by the FundaciÃ³n Galileo Galilei of the INAF (Istituto Nazionale di Astrofisica) at the Observatorio del Roque de los Muchachos. This study is part of the Large Program titled SPA Stellar Population Astrophysics: the detailed, age-resolved chemistry of the Milky Way disc (PI: L. Origlia), granted observing time with HARPS-N and GIANO-B echelle spectrographs at the TNG.}}

   \subtitle{Atmospheric parameters of members of 16 unstudied open clusters}

   \author{R. Zhang
          \inst{1,2}
          \and
 S. Lucatello\inst{2}
 \and
 A. Bragaglia\inst{3}
 \and
 R. Carrera\inst{2}
 \and
 L. Spina\inst{2}
 \and
 J. Alonso-Santiago\inst{4}
 \and
 G. Andreuzzi\inst{5,6}
 \and
 G. Casali\inst{7,8}
 \and
 E. Carretta\inst{3}
 \and
 A. Frasca\inst{4}
  \and
 X. Fu\inst{8,3}
 \and
 L. Magrini\inst{9}
 \and
 L. Origlia\inst{3}
 \and
 V. D'Orazi\inst{2}
 \and
 A. Vallenari\inst{2}
}
   \institute{Dipartimento di Fisica e Astronomia, Universita' di Padova, vicolo Osservatorio 2, 35122, Padova, Italy \\
   \email{ruyuan.zhang@studenti.unipd.it; sara.lucatello@inaf.it}
   \and
   INAF-Osservatorio Astronomico di Padova, vicolo Osservatorio 5, 35122, Padova, Italy
   \and
   INAF-Osservatorio di Astrofisica e Scienza dello Spazio di Bologna, via P. Gobetti 93/3, 40129, Bologna, Italy
  \and
  INAF-Osservatorio Astrofisico di Catania, Via S. Sofia 78, 95123, Catania, Italy 
  \and           
   Fundaci\'on Galileo Galilei - INAF, Rambla Jos\'e Ana FernÃ¡ndez P\'erez 7, 38712, BreÃ±a Baja, Tenerife, Spain
  \and
  INAF-Osservatorio Astronomico di Roma, Via Frascati 33, 00078, Monte Porzio Catone, Italy
  \and
  Dipartimento di Fisica e Astronomia, UniversitÃ  degli Studi di Firenze, Via G. Sansone 1, 50019, Sesto Fiorentino, Firenze, Italy 
  \and
  KIAA-The Kavli Institute for Astronomy and Astrophysics at Peking University, Beijing 100871, China
  \and
  INAF-Osservatorio Astrofisico di Arcetri, Largo E. Fermi, 5, 50125, Firenze, Italy
            }

   \date{Received }

  \abstract
    {Thanks to modern understanding of stellar evolution, we can accurately measure the age of Open Clusters (OCs). Given their position, they are ideal tracers  of the Galactic disc. Gaia data release 2, besides providing precise parallaxes, led to the detection of many new  clusters, opening a new era for the study of the Galactic disc. However, detailed information on the chemical abundance for OCs is necessary to accurately date them and to efficiently use them to probe the evolution of the disc.}
   {Mapping and exploring the Milky Way structure 
is the main aim of the Stellar Population Astrophysics (SPA) project. Part of this work involves the use of OCs and the derivation of their precise and accurate chemical composition.
We analyze here a sample of OCs located within about 2 kpc from the Sun, with ages from about 50 Myr to a few Gyr.
}
  {We used HARPS-N at the Telescopio Nazionale Gaileo and collected very high-resolution spectra (R = 115\,000) of  40 red giant/red clump stars in 18 OCs (16 never or scarcely studied plus two comparison clusters). We measured their radial velocities and derived the stellar parameters ($T_{\textrm{eff}}$, log g,  v$_{micro}$, and [Fe/H]) based on equivalent width measurement combined with 1D - LTE atmospherical model. }
{We discussed the relationship between metallicity and Galactocentric distance, adding literature data to our results to enlarge the sample and taking also age into account. We compared the result of observational data with that from chemo-dynamical models. These models generally reproduce the metallicity gradient well. However, at young ages we found a large dispersion in metallicity, not reproduced by models. Several possible explanations are explored, including uncertainties in the derived metallicity. We confirm the difficulties in determining parameters for young stars (age < 200 Myr), due to a combination of intrinsic factors (activity, fast rotation, magnetic fields, etc) which atmospheric models can not easily reproduce and which affect the parameters uncertainty.}
   {}


   \maketitle
%

\section{Introduction}
Most of our knowledge of stellar physics, on the ages of stars and on their evolution has been acquired thanks to the study of star clusters, of their formation and evolution and of their stellar populations.
This knowledge has very general bearing in our insight on a variety of astrophysical processes. Age dating stellar populations is necessary to investigate the formation of the Milky Way \cite[e.g.][]{Ness2016}. The interpretation of light from simple stellar systems allows the building of population synthesis models used in studying the star formation history in other galaxies.  The rate and timing of mass loss is a crucial ingredient in probing the chemical evolution and feedback processes in galaxies. For all this, stellar clusters are privileged probes and test cases.

Open Clusters (OCs hereafter), in particular, provide key insight on the Galactic disc.
These objects are made of coeval,  chemically homogeneous, and dynamically bound groups of stars born from the same molecular cloud. With metallicities 
not too far off from that of the Sun ($-0.5<$[Fe/H$]<0.5$, see e.g. \citealt{netopil16,donor20,casali19}), they are generally young (most of them are $<1$ Gyr, but there are clusters with ages as high as 
8--10\,Gyr, see e.g. \citealt{kharchenko13,cantat20}) and range greatly in size: from loose associations with just a handful of stars to super star clusters, with as many as 10$^4$ members. As they form and evolve into (or in close proximity to)  the Galactic disc, they are prone to stripping and disruption, and are in fact thought to be one of the main sources of field stars \citep[e.g.][]{lada03}. 

Stellar populations in OCs cover stars from low to high mass, and different evolutionary stages, making each cluster a snapshot of stellar evolution at a given age and composition. 
 With ages covering the entire lifespan of the thin disc, OCs trace the young, intermediate-age, and old thin disc components. 
 Age can be measured for OCs with much more accuracy than for Galactic field stars, making them the ideal tool to probe the Galaxy formation and evolution, through the age-metallicity relation, radial gradients and the comparison with theoretical models.

The Gaia results have brought a veritable revolution in our knowledge of OCs. High probability memberships based on proper motions and parallaxes have been derived for very large samples of stars, leading also to the discovery of a substantial number of new OCs \citep[see e.g.][]{Cantat2018,Castro2019,castro20,LP19,sim19}.

Presently, only a minority of OCs has been studied with high quality spectroscopic data, implying not only a lack of information on the composition for the vast majority of OCs, but also possibly inaccurate ages. In fact, a
precise metallicity is a key ingredient for the derivation of ages from photometry \citep[see e.g.][]{Bossini2019}. Moreover, the sample is likely to be affected by bias towards larger clusters, where, in the years before Gaia membership information, it was easier to successfully target actual members.

While Gaia will, at end of mission, provide distances and proper motions with a precision $<10\%$ for almost all known clusters, its spectroscopic capabilities are rather limited. The crucial third kinematic dimension (radial velocity, RV) and detailed chemical composition will need to be largely provided by ground-based complementary observations.

Recently completed and ongoing large stellar surveys, such as Gaia-ESO (GES), GALAH and APOGEE \citep[][respectively]{ges,galah,apogee}, have provided  composition and RVs for a few
thousands of stars in some hundred of OCs based on high-resolution ($\sim$20-40K) spectroscopy. This 
sample will be further increased by WEAVE \citep{Dalton}, which has the study of OCs as one of the primary goals of its Galactic Archeology Survey, and likely by 4MOST \citep{4most}.

The key feature of GES and of the future surveys resides in their ability to study multiple members of the clusters in all evolutionary phases, with tens to many hundred of stars observed in each cluster (while APOGEE, for instance, relies generally on much smaller samples).
This is in fact a crucial step in the understanding of the formation of the clusters \citep[see e.g.][]{Jeffries2014,Mapelli2015} and the evolution of the stars properties following changes in rotation, activity, surface abundances, all key constraints to modern stellar evolutionary models \citep[see e.g.][on diffusion and extra mixing]{Bertelli2018,Smiljanic2016,lagarde19}.
However, because of their spectral coverage and/or resolution, these surveys provide an incomplete chemical characterization of OC stars. An accurate determination of chemical abundances requires spectra with a very high resolution and a wide spectral coverage, to measure the full set of the Fe-peak, CNO, $\alpha$, p- and n-capture elements with high accuracy, on a par with the astrometric and photometric information provided by Gaia.

This implies that there is a need for observations that take a complementary approach, i.e. study in details with high-resolution (R $>$ 50-70000), large wavelength coverage, and high signal-to-noise spectroscopy a few stars per OCs, deriving a full chemical characterization.
 Indeed, measuring elements of all nucleosynthetic chains, which probe different nuclear reaction sites in stars, means providing robust constraints to stellar evolutionary models and to the history of the Galactic disc.

This paper is part of a series on a large project, Stellar Populations Astrophysics (SPA), at the Italian TNG telescope. In this paper we will present the atmospheric parameters and radial velocities for stars in 16 OCs never or poorly studied before with high resolution spectroscopy, plus two more as a comparison sample, while chemical abundance analysis will be presented in following papers.

In Section \ref{target} we discuss the sample selection, followed by a description of observations and of the data reduction procedure in Section \ref{data}. Determination of the spectroscopic parameters is discussed in Section \ref{parameters}, while in Section \ref{discussion} we discuss our results in comparison with outcomes from  the literature and chemo-dynamical models of the 
Galactic disc. Section \ref{conclusions}  outlines our conclusions.

    \begin{figure*}
   \centering
   \includegraphics[height=18cm,width=21cm]{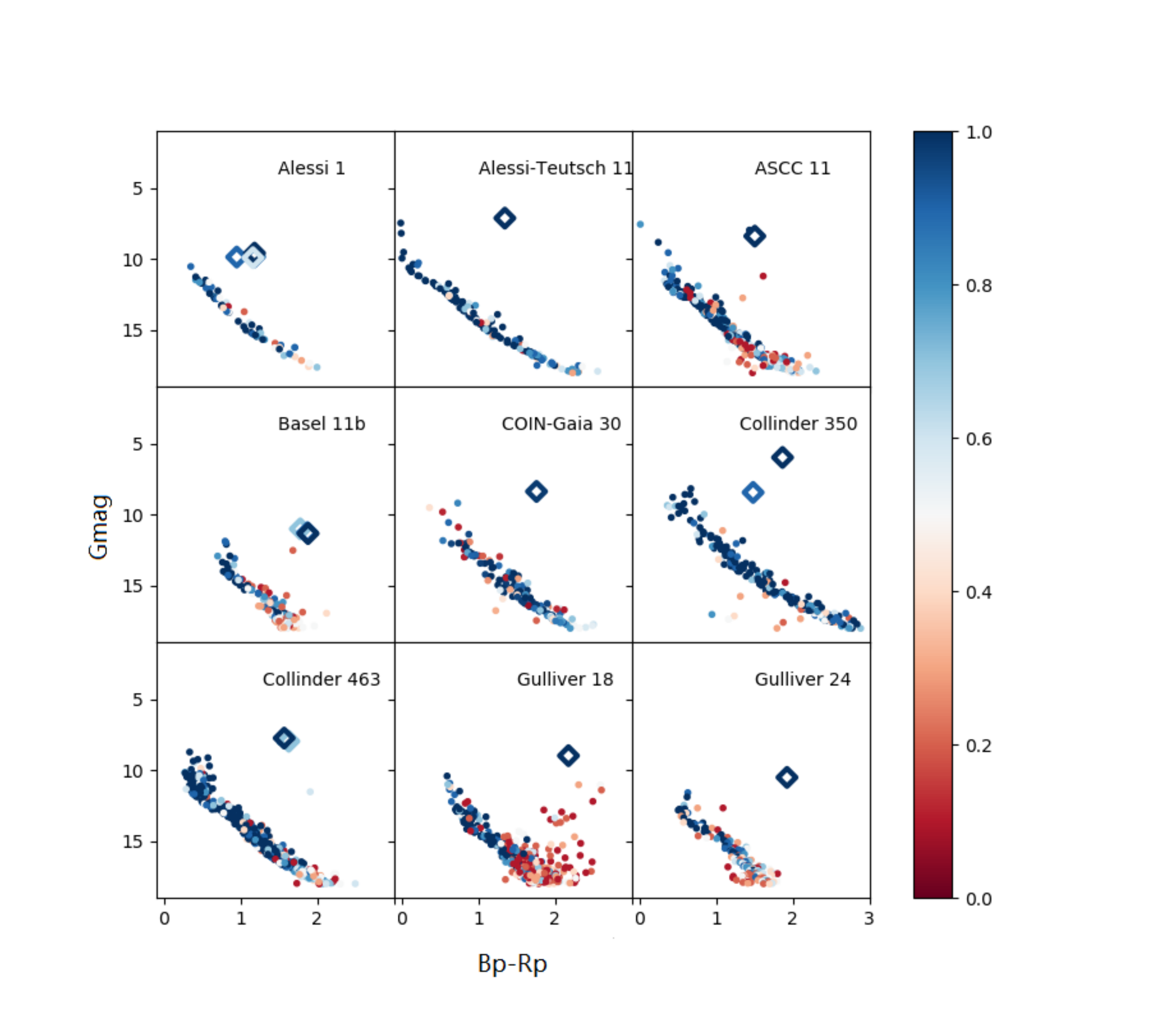}
      \caption{The CMDs of nine of the 18 open clusters; the dots are stars selected from Gaia DR2 database, the large diamonds are targets observed in this work.
      All points are coloured by membership probability \citep{Cantat2018}.}
         \label{FigCMD1}
   \end{figure*}
   
    \begin{figure*}
   \centering
   \includegraphics[height=18cm,width=21cm]{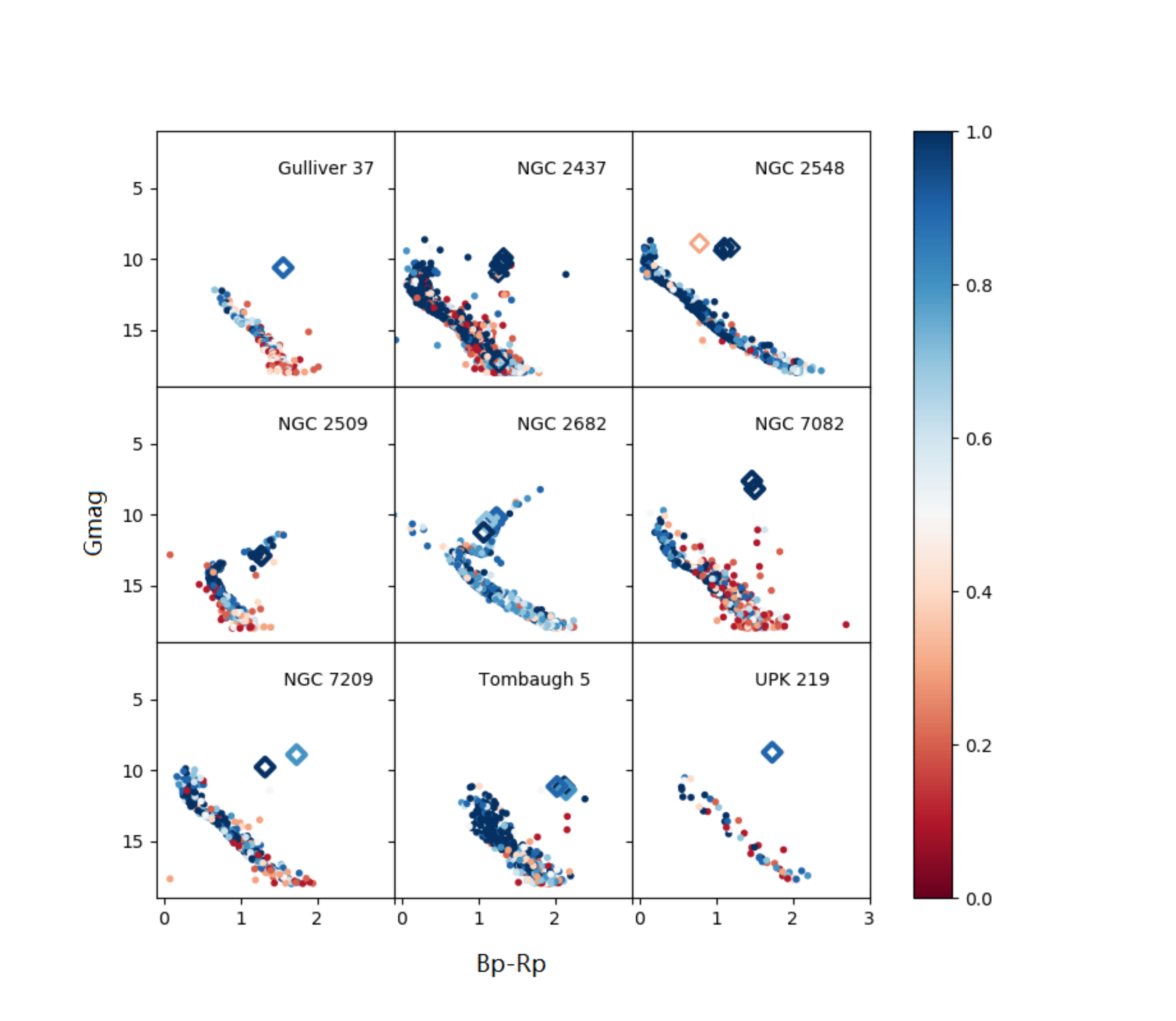}
      \caption{As in the previous figure, for the remaining nine clusters.}
         \label{FigCMD2}
   \end{figure*}

\begin{table*}[t]
\setlength{\tabcolsep}{1mm}
\begin{center}
\caption{Properties of the observed clusters. }
\begin{tabular}{l|c|c|r|r|c|c|r|r|r|r|r|r}
\hline\hline
  Cluster        &     RA     &   Dec	   &  l  &  b    &logAge & A$_V$ & plx &  pmRA & pmDE  &dist &	R$_{GC}$ &|z| \\
                 & (J2000)    & (J2000)   & (deg) & (deg) & (yr)      &  (mag)     & (mas) & (mas~yr$^{-1})$ & (mas~yr$^{-1}$)  &(pc) & (pc) & (pc)\\
\hline
\object{ASCC 11}          &03:32:13.44 &+44:51:21.6 &150.546 &-9.224 & 8.39  &0.60 & 1.141 & 0.926 &-3.030 &   867   & 9095& 139 \\
\object{Alessi 1} 	 &00:53:22.32 &+49:32:09.6 &123.255 &-13.33 & 9.16  &0.08 & 1.390 & 6.536 &-6.245 &   689   & 8726 & 159\\
\object{Alessi-Teusch 11} &20:16:30.48 &+52:03:03.6 & 87.435 & 9.268 & 8.16  &0.37 & 1.520 &-0.139 &-1.295 &   634   & 8335 &102 \\ 
\object{Basel 11b} 	 &05:58:11.28 &+21:57:54.0 &187.442 &-1.117 & 8.36  &1.56 & 0.534 & 1.046 &-4.137 &  1793   &10121 & 34   \\ 
\object{COIN-Gaia} 30 	 &01:24:19.20 &+70:34:26.4 &125.684 & 7.878 & 8.41  &1.25 & 1.346 &-6.145 & 2.067 &   767   & 8804 & 105 \\ 
\object{Collinder 463} 	 &01:48:07.44 &+71:44:16.8 &127.391 & 9.358 & 8.06  &0.79 & 1.137 &-1.715 &-0.307 &   849   & 8874 & 138  \\ 
\object{Gulliver 18} 	 &20:11:37.20 &+26:31:55.2 & 65.527 &-3.971 & 7.60  &1.59 & 0.613 &-3.198 &-5.646 &  1595   & 7816 & 110 \\
\object{Gulliver 24}	 &00:04:38.64 &+62:50:06.0 & 117.62 & 0.447 & 8.25  &1.05 & 0.636 &-3.241 &-1.57  &  1498   & 9131 & 11 \\
\object{Gulliver 37}	 &19:28:18.48 &+25:20:49.2 & 59.547 & 3.806 & 8.55  &1.33 & 0.642 &-0.775 &-3.74  &  1438   & 7712 & 95 \\
\object{NGC 2437} 	 &07:41:46.80 &-14:50:38.4 &231.889 & 4.051 & 8.48  &0.73 & 0.603 &-3.838 & 0.365 &  1511   & 9345 &106  \\ 
\object{NGC 2509} 	 &08:00:48.24 &-19:03:21.6 &237.844 & 5.840 & 9.18  &0.23 & 0.363 &-2.708 & 0.764 &  2495   & 9887 & 254 \\ 
\object{NGC 2548}	 &08:13:38.88 &-05:43:33.6 &227.842 &15.390 & 8.59  &0.15 & 1.289 &-1.313 & 1.029 &   772   & 8857 & 205 \\ 
\object{NGC 7082} 	 &21:28:44.64 &+47:06:10.8 & 91.115 &-2.859 & 7.79  &0.79 & 0.729 &-0.293 &-1.106 &  1339   & 8472 & 66 \\ 
\object{NGC 7209} 	 &22:04:53.76 &+46:30:28.8 & 95.480 &-7.296 & 8.63  &0.53 & 0.820 & 2.255 & 0.283 &  1154   & 8525 & 146 \\
\object{Tombaugh 5} 	 &03:47:56.16 &+59:04:12.0 &143.944 & 3.599 & 8.27  &2.07 & 0.561 & 0.515 &-2.388 &  1706   & 9768 & 107 \\ 
\object{UPK 219} 	 &23:27:24.96 &+65:18:36.0 &114.325 & 3.861 & 8.17  &1.20 & 1.210 &-1.734 &-2.459 &   873   & 8735 &58 \\
\multicolumn{12}{c}{Comparison clusters} \\
\object{Collinder 350} 	 &17:48:04.32 &+01:31:30.0 & 26.952 &14.773 & 8.77  &0.52 & 2.708 &4.965  &-0.019 &   371   & 8021  &94\\
\object{NGC 2682} 	 &08:51:23.04 &+11:48:50.4 &215.691 &31.921 & 9.63  &0.07 & 1.135 &10.986 &-2.964 &   889   & 8964&470  \\ 
\hline
\end{tabular}
\tablefoot{All data come from \cite{cantat20} and are based on Gaia DR2.}
\label{tab-prop}
\end{center}
\end{table*}

%


\begin{table*}
\caption{Radial velocities of targets.}
\setlength{\tabcolsep}{1.25mm}
\label{tab:RV}
\begin{tabular}{lccrrccl}
\hline\hline
   Name        &     Gaia ID  &  S/N & RV &$\sigma$RV & RV(Gaia) & $\sigma$RV(Gaia) & Notes\\
   &             & (600nm)  & km~s$^{-1}$  & km~s$^{-1}$ & km~s$^{-1}$ & km~s$^{-1}$ &\\
\hline
ASCC 11  & 241730418805573760 &93  & -13.36&0.15& -13.87  & 0.23 &\\
Alessi 1\_1 	 &402506369136008832   &93   & -5.50&0.05&-4.19&0.64  &  \\
Alessi 1\_2 	 &40250599117880752  &88   & -3.13&0.07&-5.23&0.81  &  \\
Alessi 1\_3  	&  402867593065772288  &127    & -4.57&0.20&-4.25&1.83  &  \\
Alessi 1\_4  	 & 402880684126058880    &120  & -4.29&0.03&-4.54&0.42  &  \\
Alessi-Teusch 11  &  2184332753719499904  &126   & -27.11&0.12&-27.09&0.17  &  \\
Basel 11b\_1 	 &  3424056131485038592  &125    & 2.26&0.18&3.11&0.49  &  \\
Basel 11b\_2 	 & 3424055921028900736  &86  & 1.68&0.15&5.81&1.11  &  \\
Basel 11b\_3  	 & 3424057540234289408  &66   & 2.57&0.15&2.71&1.12 &  \\
COIN-Gaia 30 & 532533682228608384   &93 & -26.66&0.13&-26.10&0.14  &  \\
Collinder 463\_1  &  534207555539397888     &143   & -9.68&0.16&-12.47&0.14  &  \\
Collinder 463\_2  & 534363067715447680& 152    & -11.64&0.12&-11.62&0.14  &  \\
Gulliver 18 	& 1836389309820904064   &87  & -1.97&0.18&-3.32&0.21  &  \\
Gulliver 24	& 430035249779499264    &78 & -30.37&0.16&-31.86&0.49  &  \\
Gulliver 37	& 2024469226291472000  &78 & -4.59&0.17&2.52&8.22 &binary? (see text)   \\
NGC 2437\_1  &3029609393042459392     &64  & 49.77&0.11&50.07&0.61  & \\
NGC 2437\_2  &  3029202711180744832  &111   &47.16&0.14&46.80&0.50  &  \\
NGC 2437\_3   &  3030364134752459904     &95  & 49.02&0.17&48.80&0.25  &  \\
NGC 2437\_4   & 3029132686034894592     &128  & 49.93&0.15& &  &  \\
NGC 2437\_5   &3029156222454419072     &50  & 49.37&0.17&49.34&0.17  &  \\
NGC 2437\_6   &3029207006148017664     &72   & 27.12&0.17&28.76&9.91  &  P$\sim3350^d$ \citep{mermilliod}   \\
NGC 2437\_7  & 3029226694277998080    &74   & 49.53&0.19&51.60&0.62  & \\
NGC 2509       & 5714209934411718784 &128    & 61.63&0.14&61.42&1.45  &  \\
NGC 2548\_1	 & 3064481400744808704 &125  & 9.06&0.14&8.68&0.29  &  \\
NGC 2548\_2	 &  3064537647636773760    &138    & 8.16&0.03&8.83&0.31  &  \\
NGC 2548\_3 	 &3064579703955646976     &77    & 8.01&0.04& 8.76&0.22  &  \\
NGC 2548\_4 	 & 3064486692144030336   &94   & 8.98&0.10&8.10 &0.67& SB  \citep{mermilliod08} \\
NGC 7082\_1 	  & 1972288740859811072   &144    & -11.89&0.15 &-11.07&0.19 &  \\
NGC 7082\_2 	  & 1972288637780285312  &151  &&&&& Double lined binary   \\
NGC 7209\_1 	  &1975004019170020736   &81   & -19.00&0.16&-18.14&0.27  &  \\
NGC 7209\_2 	  & 1975002919658397568 &161 &-17.98&0.15 & -18.89&0.35  &  \\
Tombaugh 5\_1 	  & 473266779976916480      &72    &-23.03&0.15&-22.05&1.03  &  \\
Tombaugh 5\_2 	  &473275782228263296      &54    &-22.63&0.12&-22.84 &0.26&  \\
Tombaugh 5\_3 	  &473268424940932864      &53    &-20.53&0.13&-23.45 &0.29&  \\
UPK 219 	& 2209440823287736064  & 81  & -0.07&0.15&-2.03&0.55  &  \\
\hline    
\hline 
\multicolumn{7}{c}{Comparison clusters} \\
Collinder 350\_1 	& 4372743213795720704   &240    & -13.98&0.17&-14.57&0.20  &  \\
Collinder 350\_2 	&  4372572888274176768  &70   & -16.30&0.13&-14.73&0.16  &  \\
NGC 2682\_1 	 &  604921512005266048   &141   & 33.63&0.23&34.29&0.30  &  \\
NGC 2682\_2 	 & 604920202039656064   &78   & 34.93&0.14&34.49&0.35  &  \\
NGC 2682\_3  	 & 604904950611554432 &82   & 33.93&0.20&36.83&0.96  & binary (see Tab.~\ref{tab-comp}) \\
NGC 2682\_4  	 &604917728138508160    &53   & 27.41&0.13&26.98&2.70  & binary (see Tab.~\ref{tab-comp}) \\
\hline
\end{tabular}
\end{table*}

\section {Target selection and dataset \label{target}}

In this work, we present the analysis of 40\footnote{We observed 41 stars, but one is a double-spectrum binary and was excluded from analysis.} giant stars in 18 open clusters. Sixteen of the OCs  were poorly or not at all studied before, while two of them were previously studied and work as comparison samples.
These clusters were selected because they are in the Sun vicinity so we could obtain good quality, high-resolution spectra and are old enough to have stars evolved off the main sequence.
The target stars were not selected directly from the Gaia DR2 catalog, but from the membership analysis done by \cite{Cantat2018}. We targeted only stars with high membership probability and we have from one (in 8 cases) to a maximum of 7 stars (in 1 cluster) per OC.

We selected (almost) only red clump stars, not giants in general. They are a rather homogeneous type of stars, bright enough (their absolute magnitude is $M_V \sim0-1$ mag, \citealt{girardi}) to be observed at very high-resolution in the Solar neighborhood and to relatively large distance, warm enough ($T_{\textrm{eff}}\sim4500-5700$~K, \citealt{girardi}) that they can be analysed to derive meaningful parameters and abundances. Their temperature is high enough to allow a
precise abundance analysis, as opposed to cooler, upper red giant branch stars,
where line crowding (in particular at near solar metallicities) may hamper
accurate analyses (see for instance the discussion in 
 \citealt{casali20}). 

Table~\ref{tab-prop} shows information on the clusters' position (both equatorial and Galactic coordinates) and some basic parameter, such as age, distance, reddening, etc. All values come from a single homogeneous source, i.e. \cite{cantat20}, where Gaia DR2 data are used to define candidate cluster members and derive cluster properties (see that paper for details).

All these OCs are located in the Galactic thin disc, cover the Galactocentric distance range 7.7-10 kpc, reside close to the Galactic midplane, and are in the age range from about 50 Myr to 4 Gyr. Figures~\ref{FigCMD1}, \ref{FigCMD2} show the colour-magnitude diagrams (CMD) in the Gaia system of the 18 OCs,  based on the selection of \citet{Cantat2018}; the stars are coloured with membership probability and the observed targes are indicated by larger symbols.

   \begin{figure}
   \centering
   \includegraphics[width=10cm]{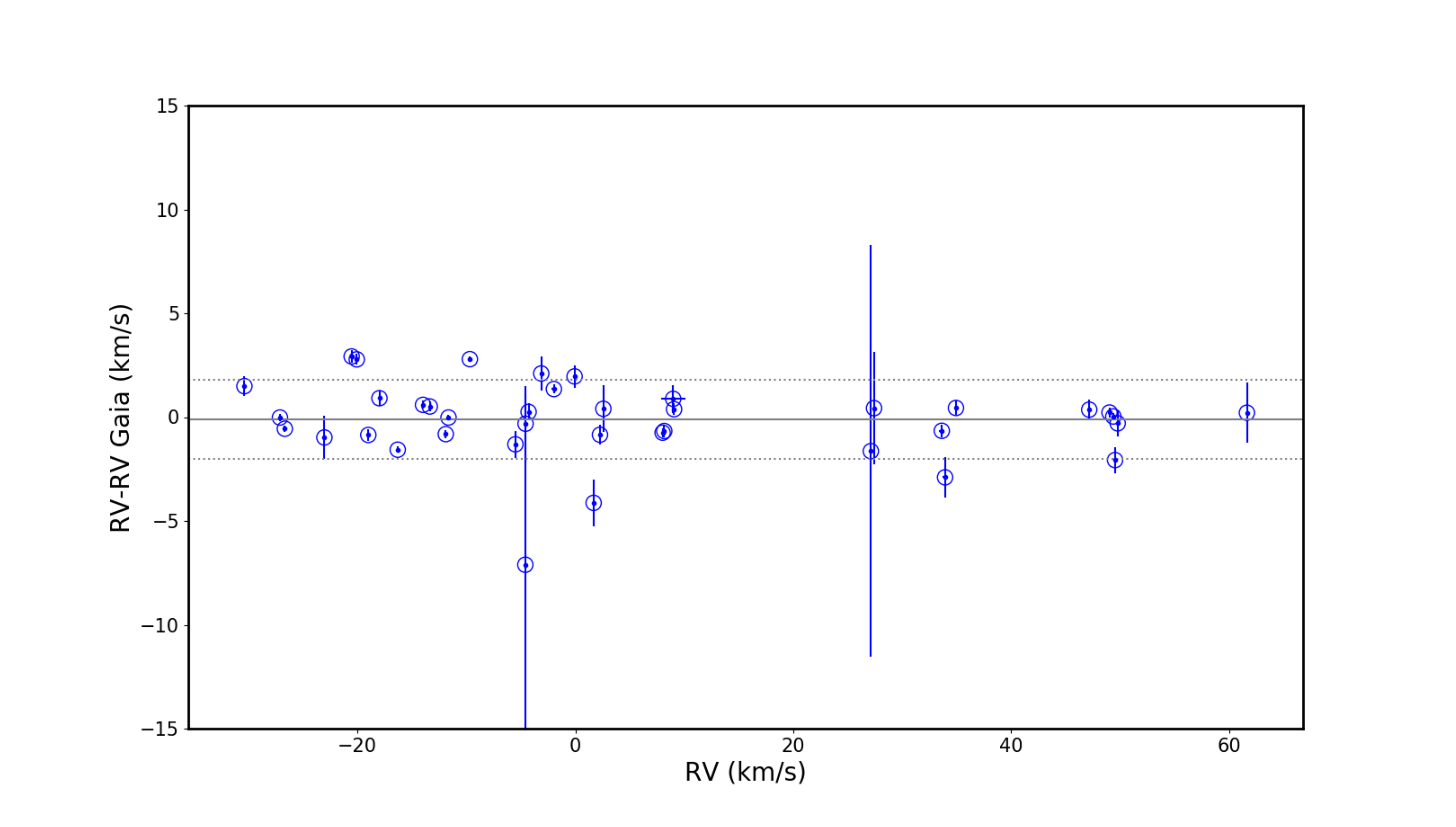}
      \caption{Difference in radial velocities for targets between Gaia and this work. The error of our sample is shown in the x-axis (it is so small that it falls within the symbol), and only the Gaia error is plotted on the y-axis, as it is much larger. The grey lines indicate the mean value of offset $-0.11 \pm 1.9$
       km~s$^{-1}$ (standard deviation).}
   \label{figRV}
   \end{figure}

\section{Observation, data reduction and radial velocities \label{data}}  

Observations were carried out in four observing runs in December 2018, January, August, and December 2019. In most cases, multiple   exposures were obtained (see Table~\ref{tab:A1} for details on the individual targets). The exposure times varied according to the star magnitude and the sky conditions; the goal was to reach at least a signal-to-noise ratio (S/N) of 50 at 500nm. We combined multi-exposure spectra  before further data analysis.

All of those data were obtained with the 3.5 m Telescopio Nazionale Galileo (TNG) at El Roque de los Muchachos Observatory (La Palma, Canary Islands, Spain). We used the two high-resolutions spectrographs HARPS-N and GIANO in GIARPS mode. This means that observation in the optical range and near-IR are executed at the same time, with the light from a single target separated using a dichroic \citep[a few more details can be found in previous SPA papers, see e.g.][]{frasca19,dorazi20}. The  analysis of GIANO data will be presented in forthcoming works; we deal here with the HARPS-N data. HARPS-N covers the wavelength range between 383 and 693 nm, with resolution R = 115000; this spectrograph is very well suited for determining radial velocity (RV) and abundances with very high precision. 

The spectra were reduced by the HARPS-N Data Reduction Pipelines (they were corrected for bias, flat field, extracted to 1d spectra, wavelength calibrated and corrected for barycentric motion) and retrieved from the Italian Center for Astronomical Archives in Trieste (\url{http://ia2-harps.oats.inaf.it/}).
Spectrum continuum normalization with cubic splines and combination were done with IRAF\footnote{IRAF is distributed by the 
National Optical Astronomy Observatory, which is operated by the Association of the Universities for Research in 
Astronomy, inc. (AURA) under cooperative agreement with the National Science Foundation.} using the tasks {\sc continuum, scombine}. 

 The calculation of RV was done on the final normalized spectra with the IRAF  task {\sc rvidlines}, which measures the wavelength shift of a list of features. For all stars,  the RV information and the error is given in Table~\ref{tab:RV}, together with the S/N measured at 600~nm. As we selected high probability members, we find very similar RVs for stars belonging to the same cluster, which further reinforces their membership.

All our stars except one (plus the binary) have an RV determined with the RVS instrument on board Gaia, which observes  the near-IR Ca~{\sc ii} triplet region at resolution R=11500. These RVs and the corresponding errors, from Gaia DR2, are also presented in Table~\ref{tab:RV}.

Figure~\ref{figRV} shows the distribution of RV offsets between the value determined in this work and that from Gaia. These offsets are generally small with an average offset of  $-0.11\pm 1.89$  km~s$^{-1}$. Owing to our small sample we have not performed further comparison. 

The few discrepant values seem probably due to the much larger errors in the Gaia RVS determinations or to binarity. Two stars in NGC~2682, namely 3 and 4 in our list, one in NGC~2437, namely 6 in our list, and one in NGC~2548, namely 4 in our list, are known binaries \citep{mermilliod,mermilliod08,geller21}.  The only Gulliver~37 star observed shows a large Gaia RV dispersion, 8.2 km~s$^{-1}$
 based on 11 transits, with an average RV of 2.5 km~s$^{-1}$, compared to our value of -4.59 km~s$^{-1}$. Moreover, we have in hand an optical high resolution spectrum for another project from which we derived a RV of $21.92\pm0.02$  km~s$^{-1}$ (Carrera et al. in prep.). Due to the large velocity dispersion reported by Gaia and the discrepancy among the different radial velocity determinations,  we consider it as a probable spectroscopic binary. {\bf All these stars do not present doubling of their lines, so the measure of the atmospheric parameters is negligibly affected by the presence of any unseen companion with respect to the precision of our analysis.}

\section{Stellar Parameters \label{parameters}}

Atmospheric parameters were determined spectroscopically, minimizing abundance trends with excitation potential, strength of the line and imposing ionisation equilibrium. In this section we describe this procedure.

\subsection{Line list and Equivalent Width measurement}
{\bf We adopted the line list for Fe~{\sc i} from \citet{Ruffoni14} and Fe~{\sc ii} from \citet{Melendez09}, selecting the lines in the 430 to 640\, nm interval.} HARPS-N spectra start from 383\, nm; however, the S/N is quite low in that region for our spectra. 
This fact, combined with the extreme crowding of features in the blue part of the spectrum, led us to discard all Fe lines bluer than 430\, nm, as their intensity could not be measured with the same accuracy of those in redder regions.

Equivalent widths  (EWs hereafter) were measured with the  code ARES, following the procedure described in \cite{2015A&A...577A..67S}, using a S/N dependent approach to set the local continuum. Visual inspection was performed for very strong lines (EW $>150$m\AA) and lines with large fitting errors (EW < 3$\times \sigma$(EW)) and we performed a manual measurement of the EW using {\it IRAF splot} when the inspection showed an issue with the ARES fitting (e.g. incorrect continuum placing due to a badly subtracted cosmic ray). Lines which turned out to be blended were discarded.

\subsection{Photometric Parameters}
The determination of the atmospheric parameters requires a set of initial parameters to be used in the first iteration of the abundance analysis determination.
For this purpose, we used photometric parameters, that are determined on the basis of observed photometric colour(s) and the latest generation of stellar isochrones.

To determine the initial stellar effective temperature ($T_{\textrm{}eff}$) we used IRFM (Infrared flux method). This method is model independent and relies on the flux on the stellar surface in infrared wavelength which relates to the surface temperature. We derived the J-K colours from 2MASS, applied the appropriate reddening for the target coordinate as derived from IRSA \citep{2011ApJ...737..103S}. The computation of photometric surface temperature follows the colour-temperature polynomial fitting based on a sample of F0-K5 type from \cite{1999A&AS..140..261A}, appropriate for stars in the range 3700-5300 K, which is consistent with our sample. 

To determine $\log g$  we used Padova CMD isochrones (PARSEC release v1.2s, \citealt{2017ApJ...835...77M} and COLIBRI, \citealt{2020MNRAS.498.3283P}, and references therein), with web interface CMD 3.4. 

For age and metallicity input for the isochrones we used information taken from the literature (see Table~\ref{tab-prop}). In case no estimate of metallicity was available, we adopted the average metallicity of the thin disc ([Fe/H] = -0.25dex \citealt{Soubiran2003}). This is admittedly low for the typical OC, however it has a negligible impact on the final parameters in Table~\ref{tab-init}.

In fact, while it is important to provide reasonably accurate input parameters in the isochrones to derive similarly reasonable photometric parameters, these values are just the starting point to build the initial model atmosphere. The actual atmospheric parameters have been determined on the basis of the observed spectra (see next section).

\begin{table}[ht]
\setlength{\tabcolsep}{1.25mm}
\begin{center}
\caption{Initial parameters from photometric data.
}
\begin{tabular}{lrrrr}
\hline\hline
  Name   & $T_{\textrm{eff}}$  & $\log$ g  &    [Fe/H] & v$_{micro}$\\
  &(K)& (dex)&(dex)  &(km~s$^{-1}$) \\
\hline
ASCC 11      &4867 & 2.36    & -0.25 &  1.0  \\
Alessi 1\_1  &4677 & 2.39    & -0.25 &  1.0  \\
Alessi 1\_2  &4600 & 2.28    & -0.25 &  1.0 \\
Alessi 1\_3  &4804 & 3.50    & -0.25 &  1.0 \\
Alessi 1\_4  &4578 & 2.25    & -0.25 &  1.0 \\
Alessi-Teusch 11 &4790 & 3.40    & 0.10 &  1.0  \\ 
Basel 11b\_1 	 &6259 & 1.70    & 0.01 &  1.0  \\ 
Basel 11b\_2 &5917 & 1.59    & 0.01 &  1.0 \\
Basel 11b\_3 &5693 & 1.83    & 0.01 &  1.0 \\
COIN-Gaia 30 &5231 & 1.70    & -0.25 &  1.0   \\ 
Collinder 463\_1 &4730 & 2.12     & -0.25 &  1.0  \\
Collinder 463\_2 &4730 &2.30     & -0.25 &  1.0 \\
Gulliver 18 &4598 & 1.50    & -0.25 &  1.0  \\
Gulliver 24	&4567 & 1.62    & -0.25 &  1.0  \\
Gulliver 37	&5095 & 1.80    & -0.25 &  1.0 \\
NGC 2437\_1 &4792& 2.24    & 0.00&  1.0 \\ 
NGC 2437\_2 &5218 & 3.38    & 0.00 &  1.0 \\
NGC 2437\_3 &5206 & 3.38    & 0.00 &  1.0 \\
NGC 2437\_4 &5087 & 2.39    & 0.00 &  1.0 \\
NGC 2437\_5 &4990 & 2.37    & 0.00 &  1.0 \\
NGC 2437\_6 &4848 & 2.29    & 0.00 &  1.0 \\
NGC 2437\_7 &4549 & 1.95    & 0.00 &  1.0 \\
NGC 2509 &4705 & 2.27    & 0.00 &  1.0 \\ 
NGC 2548\_1 &5114 & 2.50    & -0.24 &  1.0   \\
NGC 2548\_2 &5047 & 2.27    & -0.24 &  1.0 \\
NGC 2548\_3 &4853 & 2.35    & -0.24 &  1.0 \\
NGC 2548\_4 &5327 & 2.70    & -0.24 &  1.0 \\
NGC 7082 &4994 & 1.56    & -0.01 &  1.0  \\ 
NGC 7209\_1 &4799 & 2.50    & 0.01 &  1.0   \\
NGC 7209\_2 &4142 & 1.50    & 0.01 &  1.0 \\
Tombaugh 5\_1 	&5024 & 2.42    & 0.06 &  1.0  \\
Tombaugh 5\_2 &5021 & 2.40    & 0.06&  1.0 \\
Tombaugh 5\_3 &5270 & 2.44    & 0.06 &  1.0 \\
UPK 219 &5203 & 3.01    & -0.25 &  1.0    \\
\multicolumn{5}{c}{Comparison clusters} \\
Collinder 350\_1 &4200 & 1.30    & 0.10 &  1.0 \\
Collinder 350\_2 &5300 & 3.20    & 0.10 &  1.0 \\
NGC 2682\_1 &4537 & 3.61    & 0.00 &  1.0 \\ 
NGC 2682\_2 &4601 & 2.50    & 0.00 &  1.0 \\
NGC 2682\_3 &4823 & 2.90    & 0.00 &  1.0 \\
NGC 2682\_4 &4967 & 3.36    & 0.00&  1.0 \\
\hline
\end{tabular}
\label{tab-init}
\end{center}
\end{table}

\subsection{Spectroscopic Parameters}
The atmospheric parameters analysis was performed  using the code MOOG (used through the python wrapper pymoogi\footnote{https://github.com/madamow/pymoogi}, based on the MOOG 2019 version), a 1D, Local Thermodynamic Equilibrium (LTE) stellar line analysis code. 
The model atmospheres were calculated interpolating the ATLAS9 stellar atmosphere library \citep{2003IAUS..210P.A20C} in 1D plane-parallel geometry. While in principle spherical models should be better for giants, in practice this choice does not introduce significant differences \cite[see e.g.]{casali20}.

Atmospheric parameters were determined iteratively, by varying their values in the input model atmosphere until excitation and ionisation equilibria were reached.

In other words, $T_{\textrm{eff}}$'s and microturbulent velocities ($v_{micro}$) were determined by minimising the trends of Fe I abundance as determined from different lines as a function of their excitation potential and their intensity, respectively; 
surface gravity was determined by matching (within the errors) output Fe I and Fe II abundances. While we adopt the solar abundances from \cite{grevesse98}, we have also
repeated the procedure for a solar spectrum collected with HARPS-N during one of our runs, 
finding an Fe abundance essentially identical to that reported in \cite{grevesse98}.
Errors on the atmospheric parameters were derived following the same procedure outlined in \cite{2010ApJ...709..447E}. 

The resulting atmospheric parameters along with their associated uncertainties are reported in Table~\ref{tab:A2} for all individual targets. The mean value of metallicity for each cluster is quoted in Table~\ref{tab-mean} (where, for  Collinder~350, we use only the warmer star).

\begin{table}[ht]
\setlength{\tabcolsep}{1.25mm}
\begin{center}
\caption{Mean RV and metallicity for the observed clusters. 
}
\begin{tabular}{lrrrrr}
\hline\hline
  Cluster    & RV  & $\sigma$RV  &    [Fe/H] & $\sigma$[Fe/H] & N\\
  &(km~s$^{-1}$)  &(km~s$^{-1}$)  &   (dex)       & (dex) &\\
\hline
ASCC 11      &-10.94 & 0.19    & -0.14 &  0.05 & 1  \\
Alessi 1     &-4.37 & 0.85	 & 0.00 & 0.08 &4  \\
Alessi-Teusch 11 & -27.11 & 0.12 &-0.19 & 0.05  &1 \\ 
Basel 11b 	 & 2.17 & 0.37 &-0.01 & 0.05 &3  \\ 
COIN-Gaia 30 & -26.66 & 0.13  &0.03 & 0.05 &1 \\ 
Collinder 463 &-10.66 & 0.98 &-0.15 & 0.05 &2 \\ 
Gulliver 18 & -1.97 & 0.18 &-0.10 & 0.05 &1 \\
Gulliver 24	& -30.37 & 0.16  &-0.10 & 0.05 &1\\
Gulliver 37	& -4.59 $^a$& 0.17   &0.10 & 0.05 &1 \\
NGC 2437 &  49.13$^a$& 0.93	  &0.00 & 0.04 &7 \\ 
NGC 2509 & 61.63 & 0.14	  &-0.10 & 0.06&1  \\ 
NGC 2548 & 8.41$^a$ & 0.46	  &-0.02 & 0.03 &4 \\ 
NGC 7082 & -11.89 &	0.15  &-0.15 & 0.05 &1 \\ 
NGC 7209 & -18.49 &	 0.51  &-0.04& 0.04 &2 \\
Tombaugh 5 	& -21.21 & 1.30  &0.05 & 0.05 &3\\ 
UPK 219 & -0.07 & 0.15	 &0.02 & 0.05   &1\\
\multicolumn{6}{c}{Comparison clusters} \\
Collinder 350 &-15.14 &1.16 & 0.00 & 0.05  &1 \\
NGC 2682 & 34.28$^a$& 0;65	 &0.03& 0.03  &4\\ 
\hline
\end{tabular}
\tablefoot{$^a$ Binary systems (see Tab.~\ref{tab:RV}). \\
When only 1 star is observed, $\sigma$ is the error, nor the dispersion.}
\label{tab-mean}
\end{center}
\end{table}

\subsection{Comparison with literature}

Two clusters, namely Collinder~350 and NGC~2682 (M67) have been included in the present sample for comparison purpose. The former has been investigated in a previous SPA paper \citep{casali20}, which reports results obtained with a slightly different approach for the same stars, therefore allowing the mapping of possible offsets. The latter, NGC~2682, is one of the most studied OCs and represents an ideal benchmark cluster.
A few more clusters have some recently published analyses. One of them was purposely observed (Tombaugh~5) as the existing spectra were of moderate-resolution; for two more (Basel 11b, NGC~2548) the results were published after our observations had already been taken. 
Finally, we note that some other clusters
(ASCC~11, Alessi-Teusch~11, Collinder~350, NGC~2548, and NGC~2682)
were observed  within the LAMOST survey, at R=1800 \citep{zhang19}.
However, we will not discuss this last case, limiting comparisons only to spectra with at least moderate resolution (R$>10000$).

\begin{figure*}[htbp]
   \centering
   \begin{minipage}[t]{0.48\textwidth}
   \centering
   \includegraphics[height=13.6cm,width=8cm]{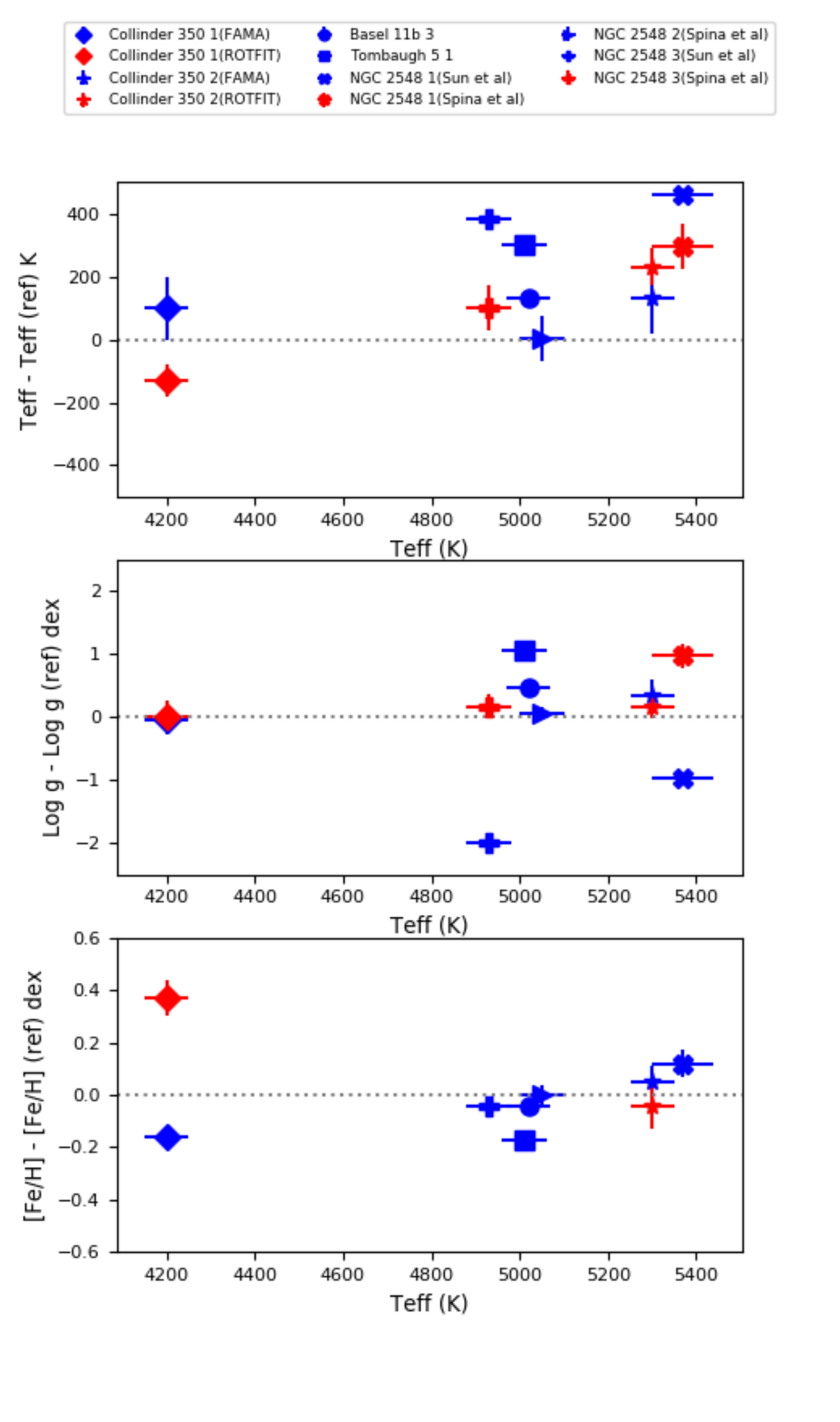}
    \end{minipage}
    \begin{minipage}[t]{0.48\textwidth}
    \includegraphics[height=13.6cm,width=8cm]{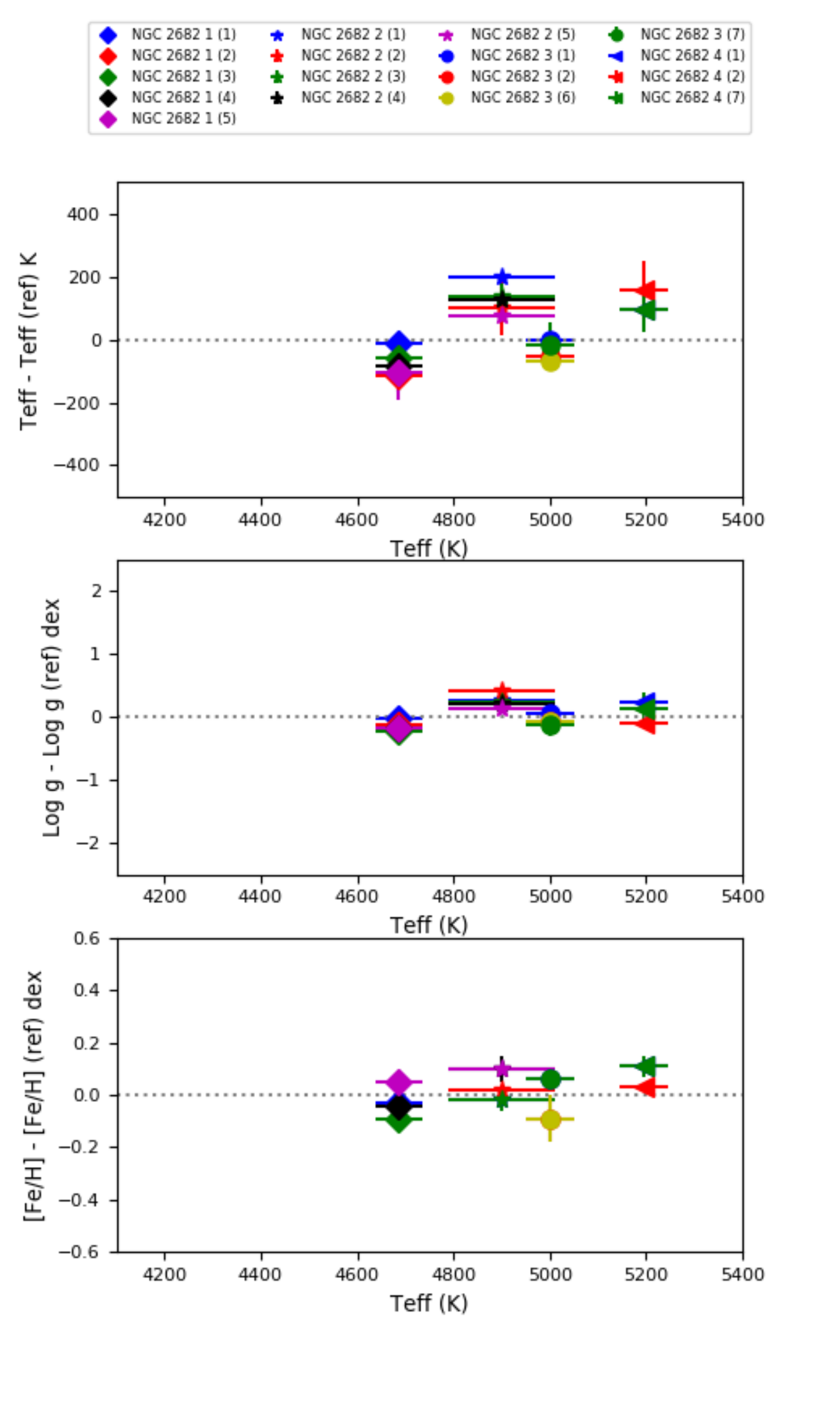}
    \end{minipage}
    \caption{Comparison of atmospheric parameters for targets with high resolution spectroscopic determinations. We plot out $T_{\textrm{eff}}$ in the x axis and the difference (our minus literature) and the error from literature on the y-axis. In the right column we show the different sources in NGC 2682 and in the left column all other  clusters. Ref: (1) \cite{jacobson11}; (2) APOGEE DR16; (3) \citep{casamiquela17} EW; (4) \cite{casamiquela17} SS; (5) \cite{gao18}; (6) \cite{luck15};(7) \cite{spina21}}
         \label{FigVibStab}
       \end{figure*}

We give details below and briefly summarise how our results compare to literature. In particular, Fig.~\ref{FigVibStab} gives a visual comparison, while Tab.~\ref{tab-comp} presents detailed values for the benchmark cluster NGC~2682 and the other stars. We remark that, for NGC~2682, there are no systematic differences among the different studies.


\paragraph{Collinder~350 -} In our sample, we collected and analysed spectra for two members of Collinder 350, one of the least studied open clusters in the solar neighborhood.
For the coolest star (4200 K) we obtained [Fe/H]=$-0.40\pm0.08$ dex, while for the hottest  (5300 K) we found [Fe/H]=$0.00\pm0.05$ dex.
In the work of \cite{2018A&A...618A..65B}, the metallicity was estimated at 0.03 dex 
from the spectrum of one single star with a resolution about  80,000. In the SPA paper by \cite{casali20}, the value of [Fe/H] was measured using two methods on the same stars as in our analysis. Even in this case the two stars show a difference in the measured metallicity, even if smaller than in the present analysis.
With ROTFIT (i.e. fitting of spectral libraries) they get $0.03\pm 0.07$ dex (cooler star) and $-0.02\pm0.09$ dex (hotter star) 
while with FAMA (i.e. automatic method based on EWs analysis) they derive $-0.24\pm0.02$ dex (cooler star) and $-0.03\pm0.06$ dex (hotter star).

 Comparing to the present paper, the results are overall in fair agreement. The hotter star shows [Fe/H]=0, a value consistent with that from both methods in Casali's work. For the cooler star, our result is more metal poor than that measured by Casali's paper using either method.   As mentioned by \citet{casali20}, we believe that at such temperature the placements of the continuum is challenging in these stars, as they are very rich in absorption lines,  leading to the less accurate metallicities for star with low temperature ($T_{\textrm{eff}}$ < 4300 K, $log$ g < 1.8 dex).

\paragraph{NGC~2682 (M67) -} This is one of the most studied open clusters and has solar metallicity and age \citep[see e.g.][]{Bertelli2018,Bossini2019}. We observed four stars, two on the red clump and single, and two red giants known to be binary systems (see Tables~\ref{tab:RV} and \ref{tab-comp}). In particular, the two single-lined binaries (stars 3, 4 in our list) were studied by  \cite{mermilliod,mermilliod08} and \cite{geller21}, who derived their orbital parameters. Star NGC~2682 3 (aka S1237) is a yellow straggler \citep{leiner16}, i.e. a star falling between the blue stragglers (BSS) and the red giants, above the subgiant branch level in optical CMDs. \cite{leiner16} used Kepler K2 data to derive a mass of 2.9$\pm$0.2 M$_\odot$, about twice the mass of turn-off stars, corroborating the notion that yellow stragglers are a later evolutionary phase of BSS.
We collected results of spectral analysis from multi-work and list them in Table~\ref{tab-comp}. The parameters are in overall good agreement with the literature.

\begin{table*}[ht]
\setlength{\tabcolsep}{1.25mm}
\begin{center}
\caption{Comparison of atmospheric parameters with literature.} 
\begin{tabular}{lccrrccrl}
\hline\hline
  star &$T_{\textrm{eff}}$&  $\sigma T_{eff}$&log g&  $\sigma$log g   &    [Fe/H] & $\sigma$[Fe/H] & Reference~~~~~~~~~~ & ~~~~~~~~~~~~Notes\\
                 &   (K) & &(dex)& &(dex)       &  & &\\
\hline
NGC 2682\_1   &4687&50&2.37&0.07&-0.05&0.05 & Present study  & RC, single\\
             &4700&    &2.40&    &-0.02&     &\cite{jacobson11}  &\\
             &4803&84&2.48&0.04&    &0.01 &  APOGEE DR16 &\\
           &4745&57&2.59&0.09&0.04&0.04&\cite{casamiquela17} EW& \\
           &4771&13&2.55&0.03&-0.04&0.05&\cite{casamiquela17} SS&\\
             &4793&    &2.55&    &-0.10&     & \cite{gao18}  &\\
\hline
NGC 2682\_2 	 &4900&110&2.76&0.10&0.02&0.05 & Present study  & RC, single \\
             &4700&    &2.50&    &0.04 &      &\cite{jacobson11}  & \\
             &4802&84&2.34&0.04&0.00 & 0.01 &APOGEE DR16 & \\
           &4762&37&2.53&0.07&0.04 & 0.04 &\cite{casamiquela17} EW & \\
           &4776&13&2.54&0.03&-0.08 & 0.05&\cite{casamiquela17} SS &\\
            &4824&    &2.62&    &-0.08 &      &\cite{gao18}  &\\
\hline
NGC 2682\_3  &5000&50&2.77&0.10&-0.03&0.03 & Present study  & SB1, $P\sim700^d$ (\citealt{geller21}),\\
            &5000&    &2.70&    &-0.09 &      &\cite{jacobson11} & yellow straggler (\citealt{leiner16})\\ 
            &5056&    &2.85&    &0.06 & 0.09 &APOGEE DR16  &\\
            &5067&    &2.85&    &0.06&0.09 
            &\cite{luck15} &\\
            &5018&71    &2.88&0.19    &-0.09 & 0.04 &\cite{spina21}  &\\
\hline
NGC 2682\_4 	  &5195&50&3.25&0.10&0.00&0.05 & Present study  &SB1, $P\sim43^d$ (\citealt{mermilliod};\\
              &5100&    &3.00&    &-0.11 &      &\cite{jacobson11} & ~~~~~~~~~~~~~~~~~~~~~~~\citealt{geller21})\\            & 5040&92&3.34&0.06&-0.03 & 0.01 &APOGEE DR16 & \\
              &5098&71    &3.12&0.26    &-0.11 & 0.04 &\cite{spina21}  &\\
\hline
Collinder 350\_1  &4200&50&1.30&0.20&-0.40&0.08 & Present study&\\
                 &4100&100&1.35&0.23&-0.24&0.01 & \cite{casali20} FAMA&\\
                 &4330&50&1.28&0.24&-0.03&0.07 & \cite{casali20} ROTFIT&\\
\hline
Collinder 350\_2  &5300&50&3.15&0.10&0.02&0.05 & Present study&\\
                 &5170&110&2.85&0.27&-0.03&0.06 & \cite{casali20} FAMA&\\
                 &5070&60&2.99&0.19&-0.02&0.09 & \cite{casali20} ROTFIT&\\
\hline
Basel 11b\_3      &4950 &50 &2.83&0.10 &-0.04&0.05 &Present study&\\
                 &4817 &86 &2.37 &0.04 &0.00 &0.01&APOGEE DR 16& 2M05581816+2158437\\
\hline
Tombaugh 5\_1     &5010&50&3.17&0.15&0.04&0.05 & Present study&\\
                 &4710& &2.10& &0.21&0.05 &\cite{2018AJ....156..244B}&\\
\hline
NGC 2548\_1      &5370&70&3.67&0.15&0.00&0.05 & Present study&\\
                &4912& &4.64&  &  &   &\cite{2020AJ....159..220S}&\\
                &5074&73&2.69&0.19&-0.12&0.05 & \cite{spina21}&\\
\hline
NGC 2548\_2      &5050&50&2.65&0.10&-0.02&0.04 & Present study&\\
               &5049&72&2.59&0.10&-0.02&0.04 & \cite{spina21}&\\
\hline
NGC 2548\_3      &4930&50&2.70&0.10&-0.01&0.05 & Present study&\\
                &4549& &4.69& & &  &\cite{2020AJ....159..220S}&\\
                &4829&72&2.53&0.19&0.03&0.04 & \cite{spina21}&\\
\hline
\end{tabular}
\tablefoot{$\sigma [Fe/H]$ is the error from metallicity measurement for present study; EW is the result obtained by the equivalent widths-based GALA software; SS is the results obtained from spectral synthesis using iSpec}
\label{tab-comp}
\end{center}
\end{table*}

\paragraph{Tombaugh~5 -} A metallicity estimate based on moderate-resolution spectroscopy also exists  for Tombaugh~5. \cite{2018AJ....156..244B} reported atmospheric parameters (from mid-res spectra, R=13 000) for five members out of seven observed. In this case two stars are in common with our sample, as they were targeted on purpose. However, only one (Tombaugh~5\_1 for us, 7701 in \citealt{2018AJ....156..244B}) is analyzed in both papers, as they considered the second (Tombaugh~5\_2 for us, 8099 in \citealt{2018AJ....156..244B}) a possible non member on the basis of its RV and previous literature. The latter star has a high probability of being a member according to \cite{2018A&A...618A..93C} and the RVs measured by us and Gaia RVS upport its membership.  In \cite{2018AJ....156..244B} temperature and gravity were derived spectroscopically and abundances were measured using EWs, with estimated errors of 0.15-0.20 dex. They found an average metallicity of 0.06 $\pm$ 0.11 dex. The mean value from our analysis is 0.05 $\pm$ 0.05 dex.
The agreement between the analyses for all atmospheric parameters is reasonable, given the large difference in resolution between the data sets (see Fig.~\ref{FigVibStab}). 

\paragraph{NGC~2548 -} A recent paper \citet{2020AJ....159..220S} is based on moderate-resolution spectra. They report the radial  and rotational velocities for nearly 300 stars in NGC~2548 (M48), measured with Hydra@WIYN spectra (R$\sim$13500) in a region about 40~nm wide,  around the Li {\sc i} 670.8~nm line.  The candidate cluster stars were selected from CMDs based on UBVRI photometry and about two thirds resulted member, both single or in binary/multiple systems, combining the spectroscopic data and Gaia DR2 results. Temperature and gravity were derived from photometric data. For a subsample of 99 well behaved, low rotational velocity, single cluster members, the value of the metallicity was derived using 16 Fe {\sc i} lines, finding an average metallicity [Fe/H]=$-0.06\pm0.007$ dex. 
This is in agreement with $-0.02 \pm 0.04$ dex from 4 stars in our work.

Two stars are in common, for which \cite{2020AJ....159..220S}  report  $T_{\textrm{eff}}$'s of 4912 K and 4549 K, with $\log g$ 4.64 and 4.69 respectively. We obtained $T_{\textrm{eff}}$'s 5370$\pm$110 K and 4930$\pm$60 K 
and our values for $\log g$ are much lower, 3.67$\pm$0.1 dex and 2.68$\pm$0.05 dex respectively, consistent with the stars being evolved rather than in the main sequence.
The gravities in \cite{2020AJ....159..220S} are derived using the Yale-Yonsei isochrones, an approach that leads to two possible solutions, corresponding to an evolved or a main sequence star. \cite{2020AJ....159..220S}  adopted the dwarf solution, and the reason behind this choice is not discussed in the paper.  
However, given the position on the CMD of the observed stars (see Fig.~\ref{FigCMD2}), we expect that the solution corresponding to an evolved (and thus brighter) star is the most appropriate one.

\paragraph{Basel~11b -} Finally,  among the 128 OCs published in \cite{donor20} and based on SDSS/APOGEE DR16, there is Basel\_11b. The paper presents the metallicity based on one single star, [Fe/H]=0. For this star we determined [Fe/H]=-0.04 and the mean iron abundance, based on 3 stars, is of $-0.013\pm 0.034$ dex. Further comparison of atmospheric parameters is found in Tab.~\ref{tab-comp}.


\section{Discussion \label{discussion}}

The average metallicities for our sample of OCs are given in Tab.~\ref{tab-mean}. These are combined with the clusters positions (in the following, Galactocentric distance and height on the plane) to discuss the metallicity  distribution in the disc, adding literature data to increase the sample and the coverage of disc properties.

\subsection{Metallicity distribution in the disc }
The study of the radial and vertical fossil gradients such as age and composition is one of the main approaches in probing the Galactic disc(s).

In this section, we discuss the observed behaviour of OCs' metallicity, age and Galactocentric distance and what information can be gained by the comparison with models for the disc. First we derive the classical radial gradient in metallicity (e.g. \citealt{friel02}), then we compare observations with models.
Our sample is distributed essentially on the Galactic plane, 
the SPA clusters are all within 0.5~kpc 
\citep{cantat20}, and generally in the solar neighbourhood, with Galactocentric distance ranging between 7.7 to $\sim$10 kpc. With one single exception -- NGC~2682, which was observed for comparison purposes -- the clusters are quite young, with 14 out of 18 clusters being younger than 0.5\,Gyr.

By combining our data with results from some selected surveys and studies, we can widen the sample to clusters with larger ages and/or distances from the Galactic centre, building a sample that allows us to probe the properties   
of the Galactic disc, and can be used to constrain models of the chemo-dynamical evolution of the disc.
  The SPA sample currently includes the present analysis as well as four more OCs in \cite{casali20}, one  from \cite{dorazi20}, and another one in \cite{frasca19}.
 We include the results from the large spectoscopic surveys, namely: a) APOGEE \footnote{We downloaded the Value Added Catalog for the Open Clusters Chemical Abundance and Mapping project (OCCAM), at https://data.sdss.org/sas/dr16/apogee/vac/apogee-occam/} \citep{donor20}, with more than 120 OCs, mostly in the 6-15 kpc range in R$_{gc}$ and with [Fe/H] from -0.5 to 0.4 dex; b) GES, covering the wide range of Galactocentric distance from 5.8 to 20 kpc, and the metallicity range -0.5 to 0.4 dex; c) and GALAH \citep{spina21}, keeping only clusters whose parameters are based on HERMES spectra and excluding those based on the recalibration of APOGEE (see \citealt{spina21} for details).
  Moreover, we include the OCCASO sample  \citep{casamiquela17}, whose OCs are located at Galactocentric distances similar to those of the SPA project (6.5 to 10.5 kpc) and with metallicities around the solar value,   ranging from $-$0.1 dex to 0.17 dex. 
  
\begin{figure*}
   \centering
   \includegraphics[height=21cm,width=15cm]{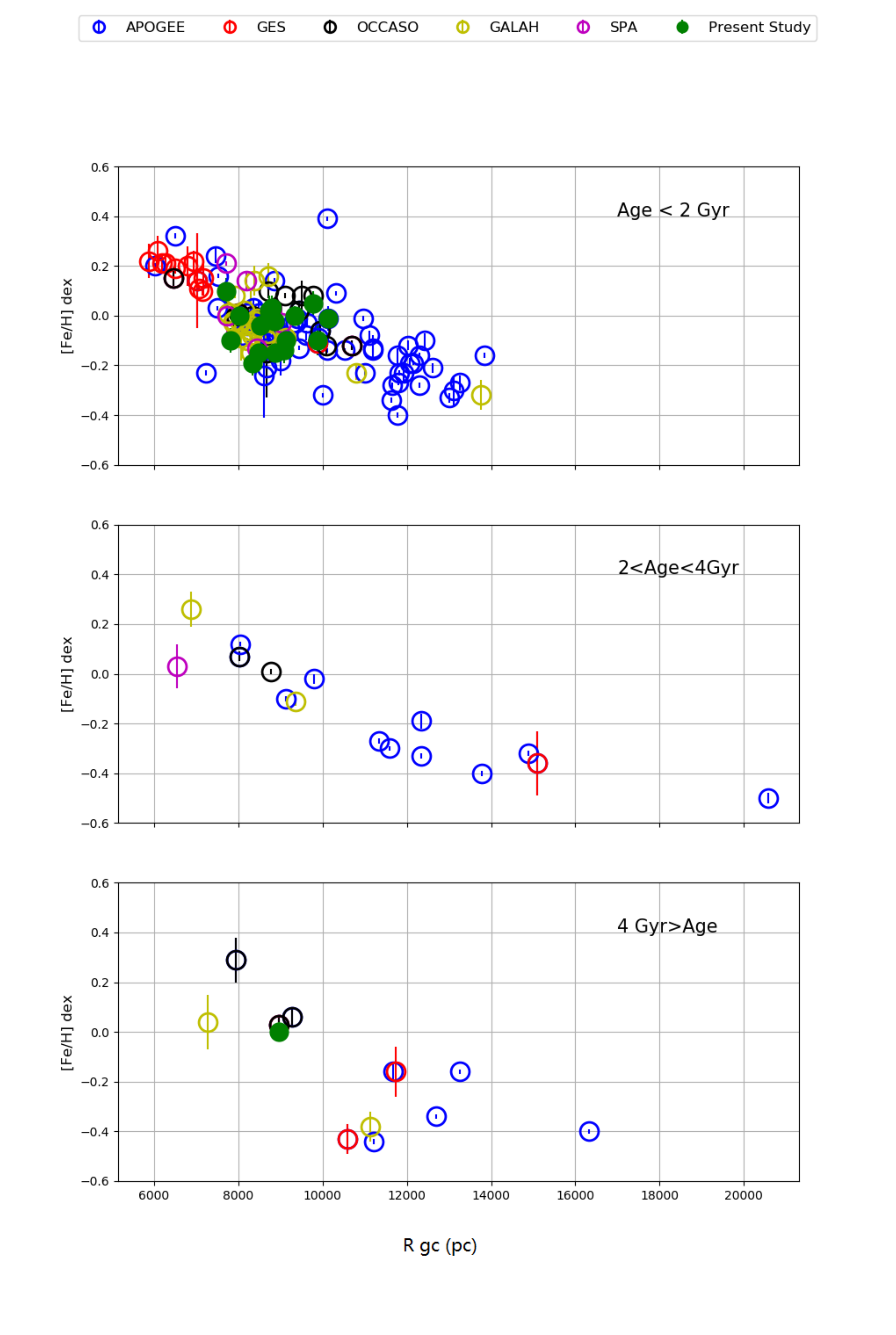}
      \caption{
      The distribution of metallicity with Galactocentric distance in three age bins (the same used in \citealt{2014A&A...572A..92M}, see Sec.~5.2). Beside our clusters, we show data from APOGEE \citep{donor20}, GES \citep{casali19}, OCCASO \citep{casamiquela17},and GALAH \citep{spina21} plus SPA results already published  \citep{frasca19,casali20,dorazi20}.}
         \label{fig_met_dist}
   \end{figure*}

 The relationship between metallicity and Galactocentric distance, classically called metallicity gradient, is shown in Fig.~\ref{fig_met_dist} for the combined sample, dividing it in three age intervals (selected because they are the same in \citealt{Minchev14b}, see below). The metallicities come from the cited sources (averaging values when more than one was available for a given cluster) and the R$_{gc}$ values come from \cite{cantat20} where R$_{gc,Sun}$=8.34 form \citet{reid14}. Table~\ref{tab-grad-noi} contains the corresponding gradients, for the whole sample and separated in age bins. 
 As the slope shows a change at about R$_{gc}$\,=14 kpc, we use that limit for inner/outer regions. The same was done in past works, as can be seen from Tab.~\ref{tab-grad}, where we list some selected literature papers. We show the values in \citet{friel02}, since this is possibly the first analysis of an homogeneous and large spectroscopic sample, albeit based on low resolution spectra, and in a few recent papers. As every work uses  different age bins and R$_{gc}$ ranges, we can make only a qualitative assessment of the results. However, the values in the two tables compare generally well; a few discrepant values are commented in the notes of Tab.~\ref{tab-grad}.
 
\begin{figure*}
   \centering
   \includegraphics[height=21cm,width=15cm]{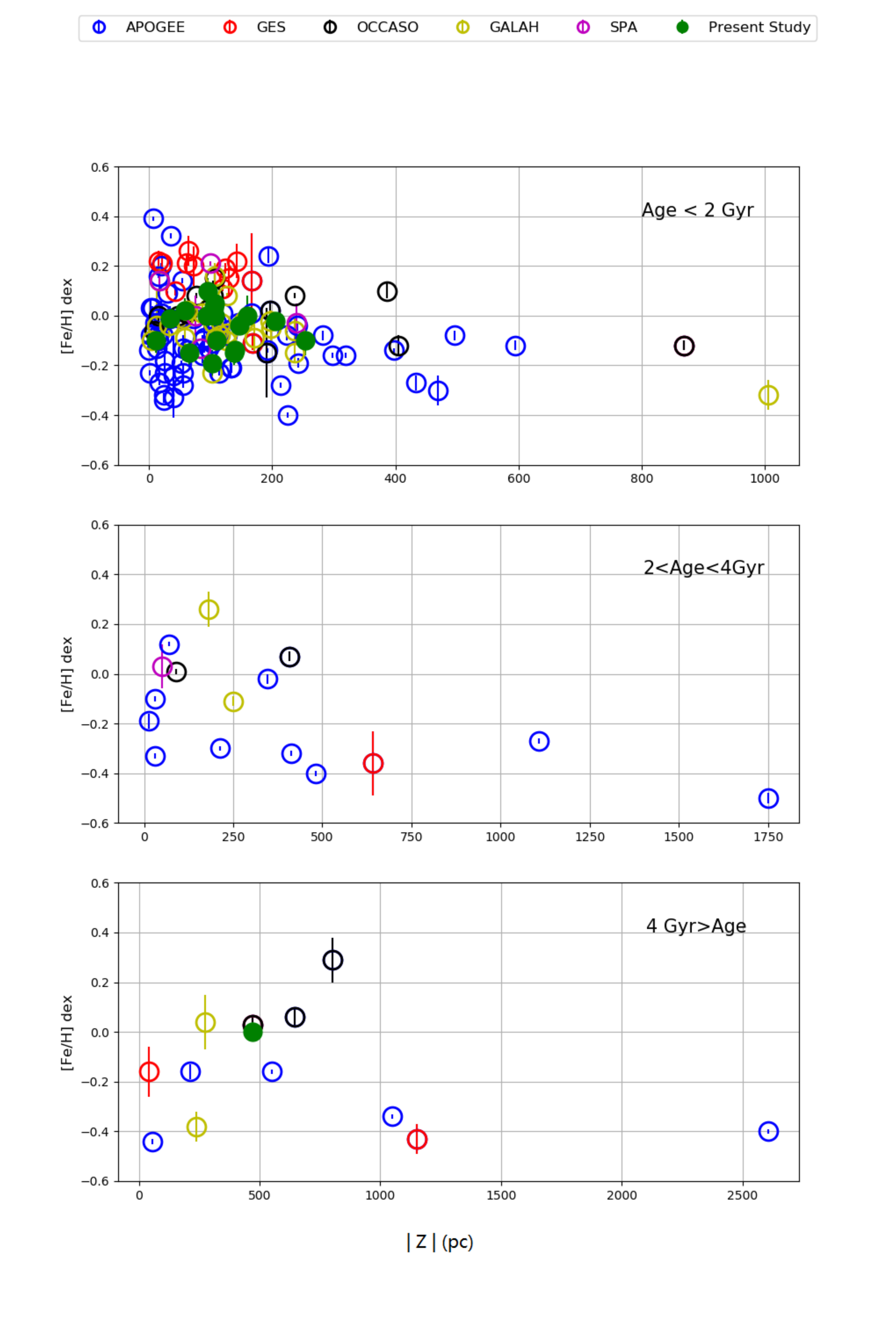}
      \caption{The distribution of metallicity with distance from mid-plate in three age bins (the same used in \citealt{2014A&A...572A..92M}, see Sec.~5.2). Beside our clusters, we show data from APOGEE \citep{donor20}, GES \citep{casali19}, OCCASO \citep{casamiquela17},and GALAH \citep{spina21} plus SPA results already published  \citep{frasca19,casali20,dorazi20}.}
         \label{fig_met_z}
   \end{figure*}

 Figure \ref{fig_met_z} shows metallicity as a function of distance from the Galactic plane, 
 divided in age bin like the previous figure. The calculated values are listed in Table~\ref{tab-grad-noi} as well. Note that, overall,  the trends with $|z|$, even in well populated bins (e.g. young clusters in the inner disk), 
 is far less statistically relevant than the corresponding slope wrt galactic radius.
 
 We also explored possible residual azimuthal gradient (only for well populated bins), finding extremely weak and moderately significant trends. For the overall sample within 14\,kpc, we found 0.0028$\pm$0.0007, while for the sample below 2\,Gyr within the same radius we found 0.0022$\pm$0.0009.  
 
 
 \begin{table}[ht]
\setlength{\tabcolsep}{1.25mm}
\begin{center}
\caption{Observed slope of the metallicity gradient.} 
\begin{tabular}{lrrrr}
\hline\hline
Age & R$_{gc}$ & d[Fe/H]/dR$_{GC}$ & d[Fe/H]/d$|z|_{GC}$&N$_{clusters}$\\
(Gyr) & (kpc) &  (dex~kpc$^{-1}$) & \\
\hline
all ages& Rgc<14&-0.066$\pm 0.005$  &$-0.249\pm0.062$ &157  \\
 all ages& Rgc>14&-0.032$\pm 0.007$ & $-0.040\pm0.028$ &4    \\
 age<2& Rcg<14& -0.059$\pm 0.006$   & $-0.236\pm0.119$ &133  \\
 2<age<4 & Rgc<14&-0.089$\pm 0.007$ & $-0.195\pm0.144$ &13   \\
 age>4& Rgc<14& 0.008$\pm0.041$     &$0.081\pm0.117$ &11   \\
\hline
\end{tabular}
\label{tab-grad-noi}
\end{center}
\end{table}
 


\subsection{Comparison with chemo-dynamical models}
It must be kept in mind that the chemo-dynamical evolution of the Galactic disc(s) is a rather complex process and the expectation is that much of the initial information will be diluted through the dynamical evolution and radial mixing of the disc.
This is expected to affect differently populations of different ages: broadly speaking, the older a cluster is, the more it has undergone dynamical evolution and radial mixing, even if other factors come into play, like the details of the formation environment and more in general the shape and mass of the disc at 
the time of formation and early evolution.

Recent models for the Galactic thin disc, e.g. \citet{Minchev13,Minchev14a,Minchev14b}, taking into account sophisticated simulations in the cosmological context \citep{martig09,martig12} along with a detailed knowledge of the chemical evolution of the disc have been able to generate theoretical gradients to compare with the observations at different ages.
However, uncertainties exist, also due to model assumptions (e.g. initial gradient of chemical abundance), especially with regard to the oldest clusters and those furthest from the Galactic plane, which have undergone a large amound of dynamical evolution during their lives.

\begin{figure*}[htbp]
\centering
\begin{minipage}[t]{1\textwidth}
\centering
\includegraphics[height=10cm,width=19cm]{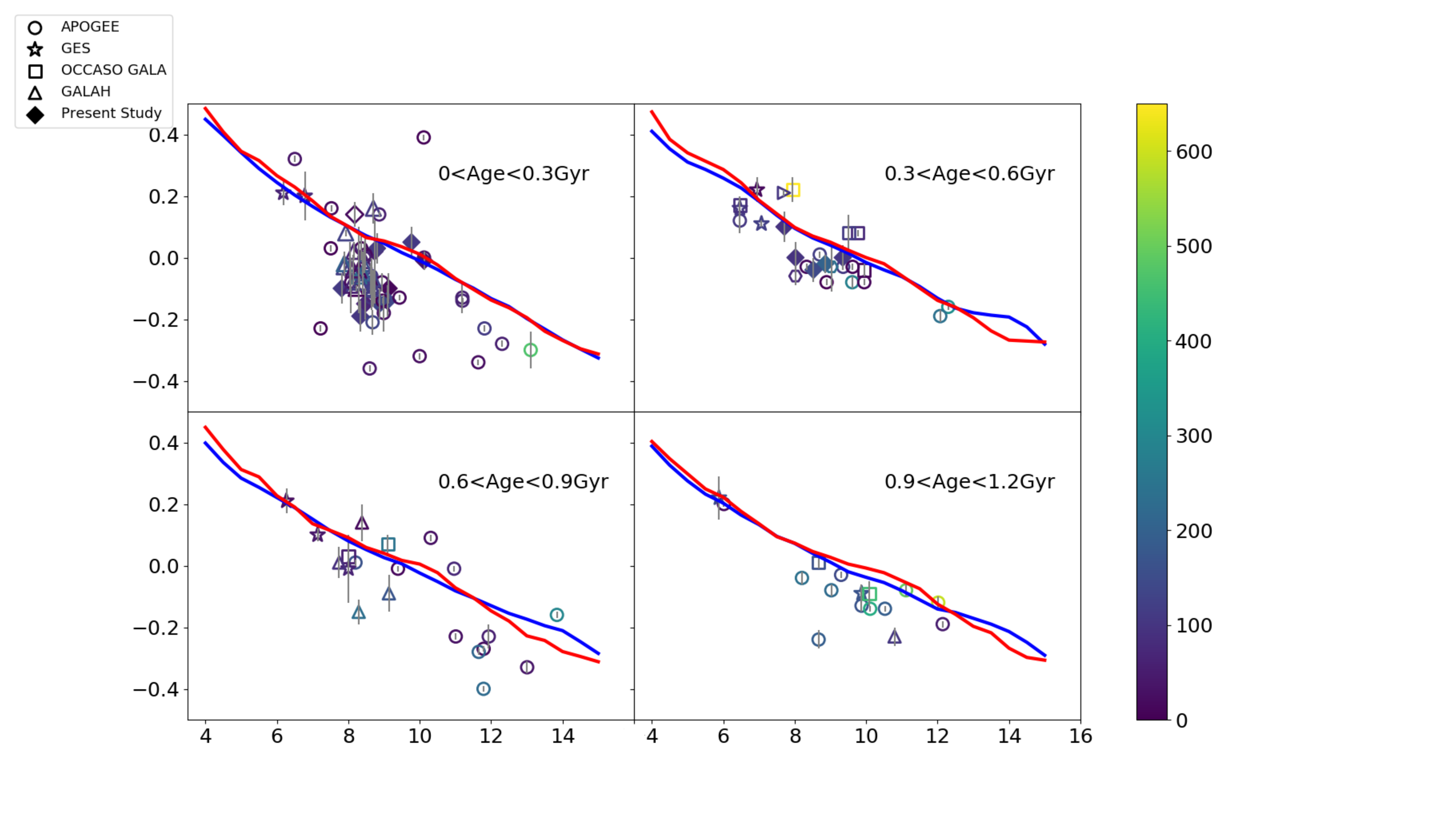}
\end{minipage}
\begin{minipage}[t]{1\textwidth}
\centering
\includegraphics[height=6cm,width=9.5cm]{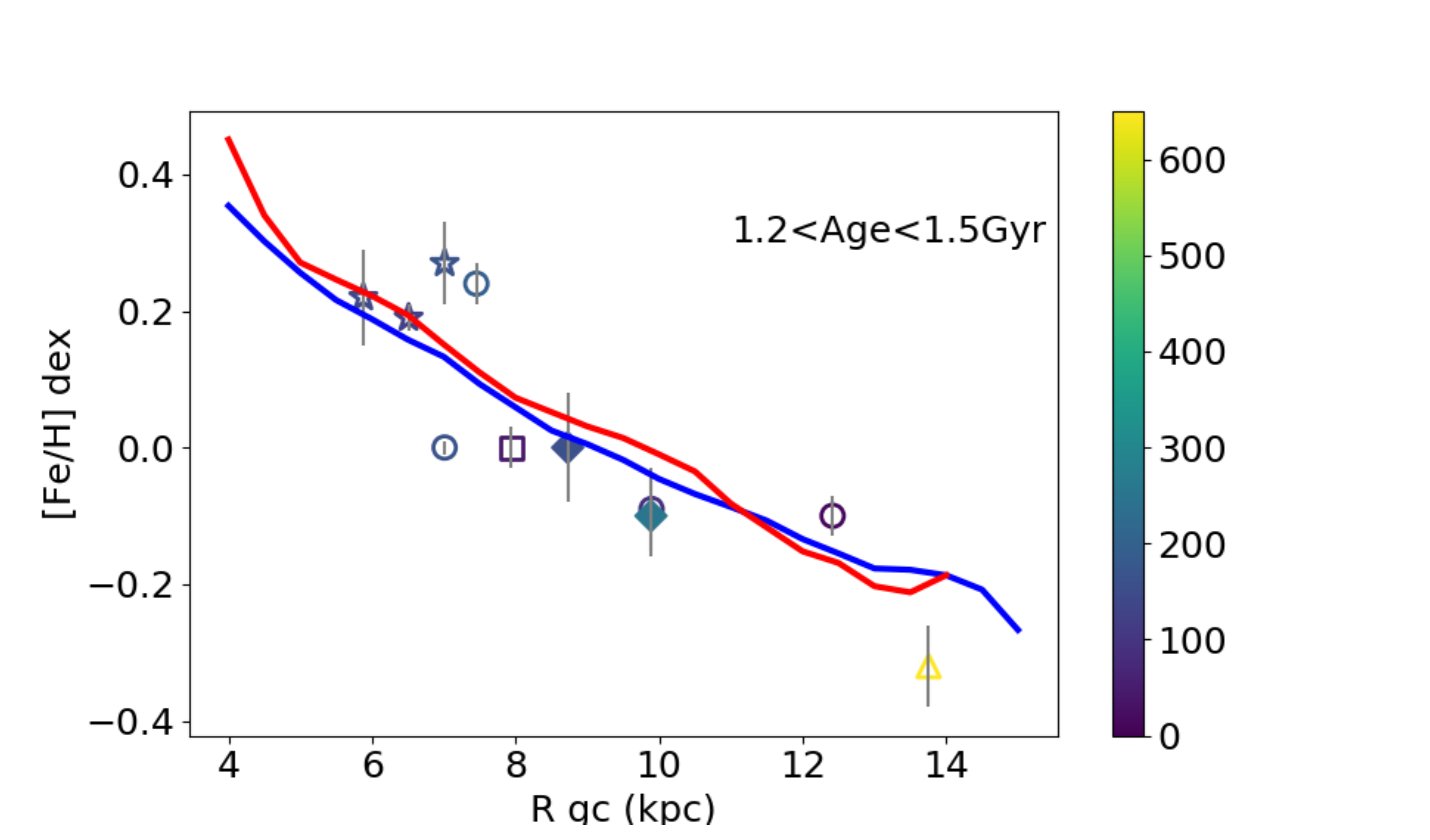}
\caption{Comparison between model predictions \citep{Minchev14a,Minchev14b} and observation for young clusters. The red and the blue lines are predictions from the MCM models for $|z| < 0.3$\,kpc and 0.3 $< |z| < $0.8 kpc respectively. The colours in the symbol indicates the distance from the Galactic plane. All considered clusters are within 0.6 kpc from the Galactic plane, with SPA clusters being all within 0.5\,kpc. The open 'diamond' in the first panel is ASCC 123 \citep{frasca19} and the 'triangle' symbol in 0.3-0.6 Gyr range is NGC 2632 \citep{dorazi20} 
The fit is generally good for clusters older than 0.3 Gyr, but the predictions fail to reproduce the data among the very young clusters (see text).}
\label{fig_metrgc}
\end{minipage}
\end{figure*}

\citet{Minchev14a} provide gradients calculated for different age intervals and ranges of distance from the Galactic plane. 
Figure~\ref{fig_metrgc} shows the comparison between the combined sample predictions from chemo-dynamical thin-disc model by \citep[][the so-called MCM model]{2014A&A...572A..92M}.
These models are calculated taking into account cosmological and dynamical properties, and chemical evolution, combining the effects of migration, the distance above the disc midplane, and then extending the model beyond the solar neighborhood. Furthermore, the uncertainties of observation, and the evolution of Galactic disc was considered.
From the MCM 
models we consider the range  0-0.8 kpc in terms of distance from the Galactic plane, which is appropriate for clusters in SPA, but also for the vast majority of the literature sample. We plot the predictions for both $|z|<0.3$\,kpc and for 0.3\,kpc$<|z|<$0.8\,kpc.
 
\begin{figure*}
   \centering
   \includegraphics[height=13cm,width=20cm]{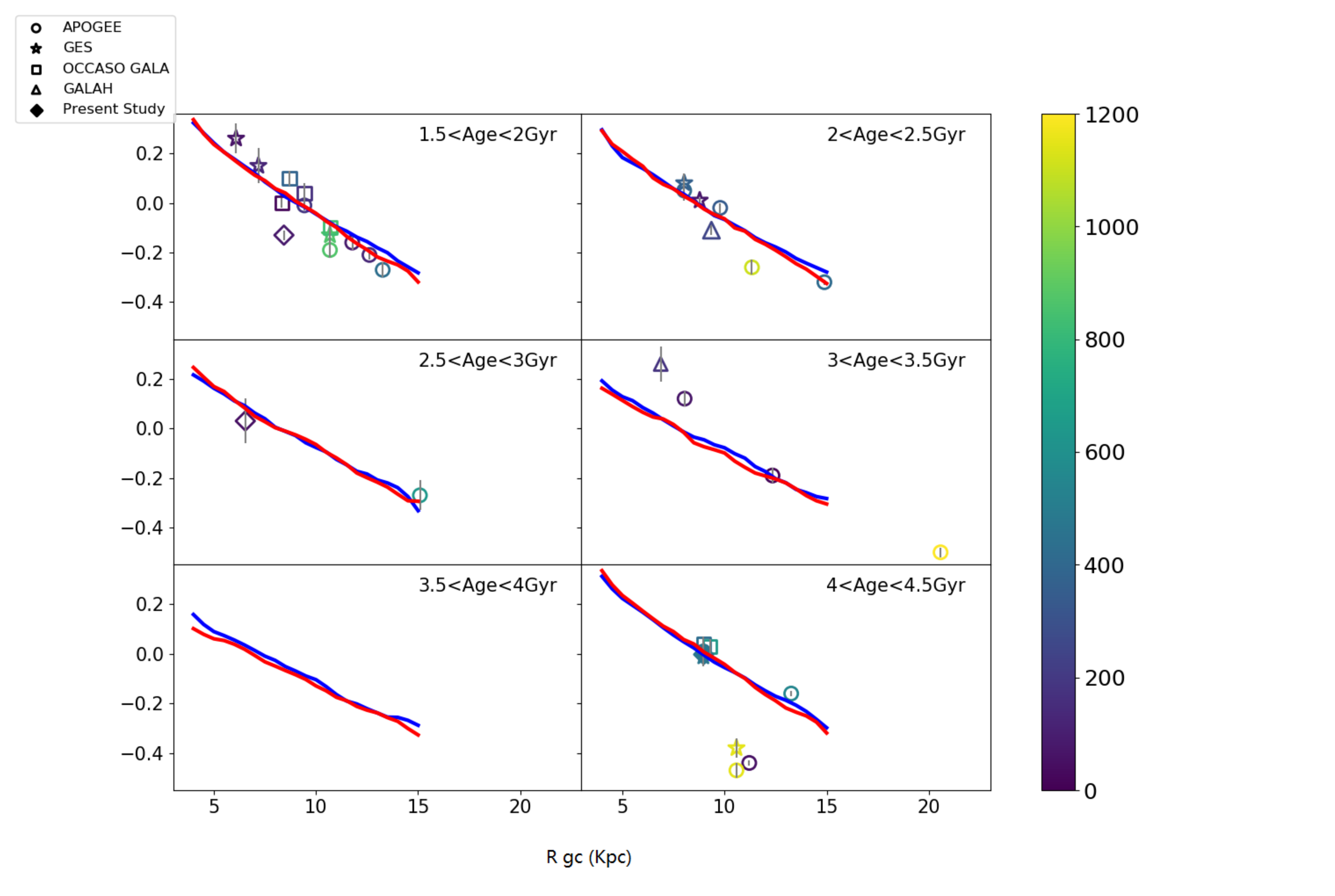}
      \caption{As in Fig.~\ref{fig_metrgc} for clusters bewteen 1.5 and 4.5 Gyr. Note the paucity of OCs older than 2.5 Gyr (in particular, only one SPA cluster is present, Ruprecth~171, from \citealt{casali20}). The data are quite well reproduced by the models; the exceptions are the old and very metal poor clusters which have $|z|$ > 1\,kpc,  further away from the Galactic plane than the plotted models.} 
         \label{fig_metrgcold}
   \end{figure*}

Figures~\ref{fig_metrgc} and \ref{fig_metrgcold}  show the comparison between MCM predictions and results from observation data, grouped according to age in bins of 0.3 and 0.5 Gyr, respectively, up to 4.5\,Gyr. 
Predictions are generally a good match for the observation in the age range from 0.3 Gyr to 4.5 Gyr.
It is worth noting that in the oldest age bin, two of the three discrepant clusters have |z|$>1$\,kpc, larger than the range the models have been calculated for, while the third has a metallicity based on one single star.
On the other hand, in the youngest bin the fitting is very poor, even if the clusters lie closer than 0.5\,kpc from the Galactic plane; we will come back to this problem later.
For a more quantitative comparison with the \citet{Minchev14b} results
we derived the gradients in our sample following the bins in age and Z in their Table 1, considered the error of metallicities in the fitting process. 
We employed their same separations in R$_{gc}$ and [z] and computed the corresponding gradients for the observed sample.
Interestingly, the gradient computed for age younger than 2 Gyr and |z| < 0.25 kpc, i.e. the bin comprising the vast majority of our sample (97 OCs), is comparable to that from the prediction:  $-0.066\pm0.008$ versus -0.057 for observed and predicted slope, respectively (see below, however, for a caveat on the very young clusters). Values are instead different in the other bins, which are, however, very scarcely populated, so it is difficult to assess if  the discrepancy is important.

Coming back to the very different distribution of OC metallicity for young ages (already noticed in past work, e.g. \citealt{spinages}), 
several possible factors might be the source(s) of the poor fit. On the observational side, inaccurate measurements of the parameters could cause the issue. For example,  a systematic underestimate in the Galactic distances would lead to outer disc clusters being compared to solar neighborhood predictions. A systematic underestimate in Z, on the other hand, would mean that we would be using predictions unsuitable for the sample.
  R$_{gc}$ and Z  are based on Gaia DR2 data, with typical uncertainties of around 5\% to 10\% for distance \citep{cantat20}, which is not large enough to generate the observed effect. Moreover, if such a systematic errors existed, they would very likely affect similarly clusters in the other age bins, a phenomenon of which there is no evidence. 
  We note that a systematic error in the adopted age could not in any way explain the poor fit: the discrepant clusters with $<0.3$\,Gyr would not be reproduced by predictions of any of the plotted ages. 
 
  Another possibility is that (part of) the metallicities we are using in the plot are not enough accurate. In particular we have checked in more detail APOGEE cluster from \citet{donor20}. Firstly, we have detected that for several cluster there are significant difference between the two latest data releases, DR16 \citep{donor20} and DR14 \citep{donor18,carrera19}, even if the same stars are used in both cases. This is for instance the case of King~7, with [Fe/H]=-0.13$\pm$0.05 and -0.04$\pm$0.01 in DR16 and DR14, respectively. These differences may be explained by the different methodologies used in the two releases to obtain the final abundances \citep[see][for a detailed explanation]{jonsson18,jonsson20}. For several clusters the values are based only on a single star with either low astrometric membership probability from \citet{2018A&A...618A..93C}, e.g. Berkeley~79, Czernik~23,  FSR~852, King~12, NGC~1857 and/or with discrepant radial velocities in comparison with the literature, e.g. Czernik~23, NGC~2311, NGC~6383. Additionally, the star observed in NGC~2311 lies well out of the cluster sequence and the star in NGC~6383 has a high rotation velocity ($>$20 km~s$^{-1}$) which complicates its analysis. In principle the APOGEE sample includes two stars in NGC~2232 with giants atmospheric parameters but this cluster does not have giants. In fact, they have a negligible astrometric membership probability from \citet{2018A&A...618A..93C} and their radial velocities ($\sim$82 km~s$^{-1}$) are in disagreement with the average value derived from 19 stars ($\sim$25 km~s$^{-1}$) with very high astrometric probabilities from Gaia DR2 by \citet{soubiran2018}. In the case of the star forming region NGC~1977, the DR16 metallicity [Fe/H]=-0.21 dex, derived from 3 stars (there is no determination in DR14), is in disagreement with the value reported in the high quality high resolution spectroscopy by \citet{netopil16} of [Fe/H]=-0.06 dex from 2 stars obtained from \citet{cunha1995}. In the case of NGC~2264, again without astrometric membership probabilities available there is good agreement between APOGEE DR16 ([Fe/H]=-0.18 dex from 23 stars) and the value determined by \citet{king2000} ([Fe/H]=-0.18 dex from 3 stars) but they disagree with the recent determination by \citet{baratella20} of [Fe/H]=+0.11 dex from a single star observed with UVES as part of Gaia-ESO. \citet{baratella20} amply discusses the problems of analysing spectra of MS stars in young stars and tries to devise a more robust method. In contrast, there are other cases, such as Berkeley~33, NGC~136, and SAI~116, whose metallicities seem reliable although they are based only on 4, 1 and 2 stars, respectively. These stars have a very high astrometric membership probability from \citet{2018A&A...618A..93C} and the derived radial velocities are in good agreement with other values reported in the literature.
  
\begin{figure*}
   \centering
   \includegraphics[height=6.5cm,width=20cm]{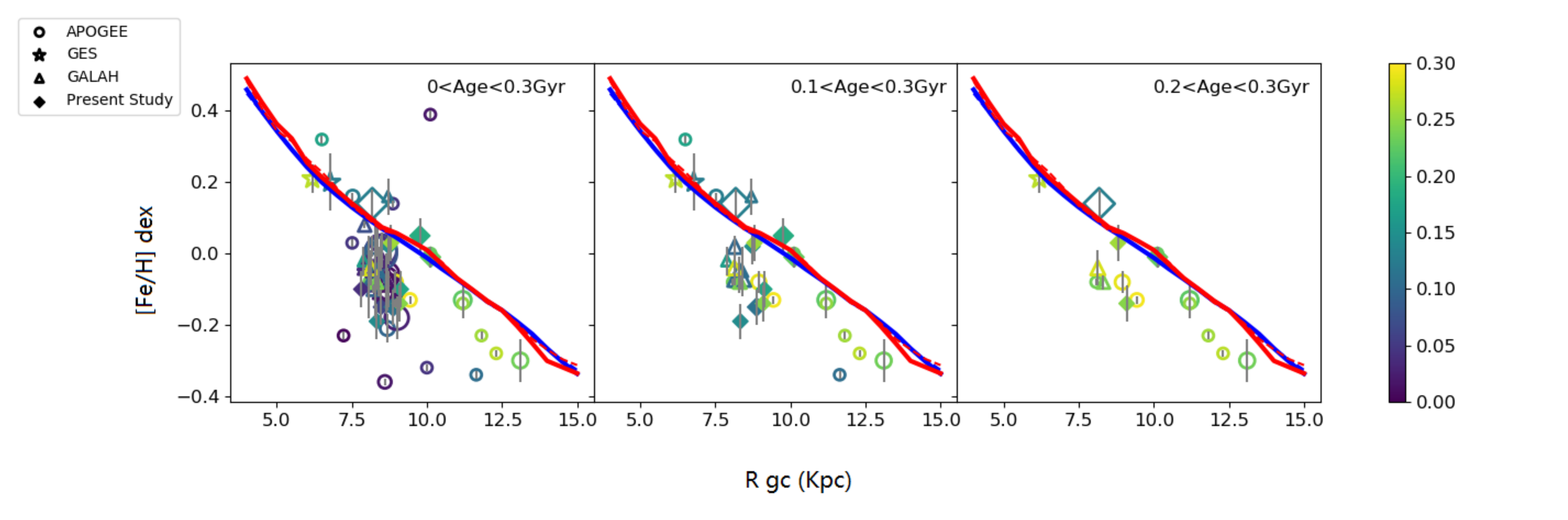}
      \caption{As in Fig 5, but colour coding the symbols by their age.The size of the symbols is proportional to the number of stars studied in the cluster to obtain the mean value reported here. (The open diamond represents ASCC 123 from \cite{frasca19}). The left panel shows all the clusters younger than 300Myr, while the middle and right panels show the fit without clusters younger than 100Myr and 200Myr, respectively. The fit to the models is quite reasonable when considering clusters older than 200Myr.   }
         \label{fig_metrgcyoung}
   \end{figure*}

 
 All these cases seem to indicate that we have to take the metallicity of young clusters with care. We also tried to understand which ages are more problematic.
 We show in Fig.~\ref{fig_metrgcyoung} only clusters younger than 300 Myr, with the symbol size proportional to the number of stars available in each cluster. Apparently, the sample size is not the (main) source of the problem.
 
 The clusters showing the larger discrepancy with respect  to the model are younger than 200 Myr, and especially  younger than 100 Myr. This is affecting analyses done both in the optical and the IR. As already discussed in literature \citep{yana19, baratella2020, spina20}, in young dwarves chromospheric effects and considerable magnetic fields are at play, making a traditional 1-D LTE analysis based on minimising trends for Fe lines less than optimal. In fact, an intensification of strong absorption lines, those forming near the top of the stellar photosphere where the magnetic fields are more vigorous, has been observed as a function of the activity level during the stellar cycle \citep{yana19, spina20}. The cause of this effect can be imputed to  Zeeman broadening of atomic lines or the effect of cool stellar spots. However, it is possible that many other phenomena related to the chromospheric activity, which are neglected in stellar models, are simultaneously at work contributing to this spectral variability. Interestingly, these problems for young stars have essentially only been studied and discussed in the case of MS or PMS stars. Our sample allow to shows clearly for the first time that the effect(s) extend also to giants. Indeed, this is not surprising, as the challenges posed by the modeling of atmospheres and spectra for giants are even more severe than those for dwarfs.\\

\section {Summary  \label{conclusions}}
This paper provides the atmospherical parameters for 40 red giants/red clumps in 18 open clusters covering the Galactocentric distance range 7.7 < R$_{gc}$ <10 kpc. Almost all of them are young clusters, with 15 OCs  between 40 and 600 Myr, two around 1.5 Gyr,  and one (NGC~2682) at 4.2 Gyr. Their parameters were measured using very high-resolution, high S/N HARPS-N spectra,  the EW method and 1D-LTE atmosphere models. The main results comprise:

(i)  Very precise radial velocities 
with uncertainties of 0.05-0.25  km~s$^{-1}$ were measured. The offset between our results and that from Gaia DR2 is  -0.11$\pm$ 1.9 km~s$^{-1}$.
 
 (ii) Accurate stellar parameters were derived. For the stars located in the effective temperature range of 4200-5800 K uncertainties in $T_{\textrm{eff}}$ are around 60 K, and  0.12 dex in surface gravity. All of the clusters have metallicity close to solar, with a deviation within 0.05 dex.
 
 (iii) We compared five of our clusters (Collinder~350, NGC~2682, Tombaugh~5, NGC~2548, Basel~11b) with results from previous work, finding no systematic bias in our determination. A few discrepant cases were examined.
 
 (iv) We explored the trend between metallicities and Galactocentric distance 
 combining our data with clusters from APOGEE, GES, OCCASO, GALAH, and other SPA OCs (ASCC~123, Gulliver~51, Ruprecht~171, NGC~2632, NGC~7044). 
 We confirmed the variation of slope in the metallicity gradient near R$_{gc}$=14 kpc
 and found a gradient slope for the inner disc similar to other literature works.
 
 (v) We used the combined sample to 
 compare to the chemo-dynamical predictions
 by \citet{Minchev14a}, finding good agreement for clusters older than 0.3 Gyr.
 Conversely, younger clusters show a large dispersion, not predicted by models, with observed metallicities too low for the Galactocentric position (not explainable by radial mixing, given the young cluster age).

 (vi) We examined the cases of young clusters with low metallicity, finding that some results may be doubtful. However, there seems to be a general difficulty in deriving accurate metallicities for clusters younger than about 200 Myr, both for dwarfs (confirming literature works) and giants. We tentatively ascribe this to
 the impropriety of the traditional analysis for young stars, as discussed in the previous section.
 
 

\begin{table*}[ht]
\setlength{\tabcolsep}{1.25mm}
\begin{center}
\caption{Slope of the metallicity gradient from selected literature papers.
}
\begin{tabular}{llccl}
\hline\hline
\multicolumn{1}{c}{Reference}&\multicolumn{1}{c}{sample}&\multicolumn{1}{c}{d[Fe/H]/d R$_{gc}$}&\multicolumn{1}{c}{N OCs} &\multicolumn{1}{c}{Comment}\\
\hline
\cite{friel02}  &low-res optical spectra &-0.059$\pm 0.010$&39 &7<R$_{gc}$<16 kpc, all ages$^a$ \\
\cite{reddy16} &high-res optical spectra &-0.052$\pm 0.011$&67 &6<R$_{gc}$<12 kpc and $|z|<$500 pc\\
               &&-0.015$\pm 0.007$&12 &12<R$_{gc}$>24 kpc$^b$ \\
\cite{carrera19} &APOGEE DR14, GALAH DR2&-0.052$\pm 0.003$&46 &6<R$_{gc}$<13 kpc$^c$ \\
                 &&-0.077$\pm 0.007$&  &6<R$_{gc}$<11kpc$^c$ \\
                 &&0.018$\pm 0.009$&  &11<R$_{gc}$<13 kpc$^c$ \\
\cite{casamiquela19} &high-res optical spectra &-0.056$\pm 0.011$&18 &R$_{gc}$<12 kpc, all ages$^d$ \\
\cite{donor20} &APOGEE DR16&-0.068$\pm 0.004$&68&R$_{gc}$< 13.9 kpc, all ages$^e$  \\
               &&-0.009$\pm 0.011$&3&R$_{gc}$>13.9  kpc, all ages\\
\hline
\end{tabular}
\tablefoot{$^a$ the slope is also given in different age ranges, with a gradient steepening for increased age: -0.023, -0.053, -0.075 for age $<$2, 2-4, and $>$4 Gyr, respectively (a similar result was found by \cite{andreuzzi11} using [Fe/H] from high resolution spectroscopy, who found values of -0.07 and -0.15 for age lower and larger than 4 Gyr, respectively). \\ 
$^b$ They do not consider it representative of the disc midplane (their Sec.~5)\\
$^c$ The first value is obtained for OCs where at least 4 stars were measured, the second and third to the whole sample. The gradient in the outer bin becomes -0.04$\pm$0.01 if the two low metallicity OCs close to R$_{gc}$11 kpc are excluded.\\
$^d$ No significant difference in the age ranges covered. \\
$^e$ Also divided for age: -0.50 (13 OCs, age$<$0.4 Gyr), -0.073 (16 OCs, 0.4-9,7 Gyr), -0.066 (27 OCs, 0.8-2 Gyr), and -0.094 (12 OCs, $>$2 Gyr).\\}
\label{tab-grad}
\end{center}
\end{table*}


 \begin{acknowledgements} 
We thank the TNG personnel for help during the observations and I. Minchev for sharing his evolutionary models. This research used the facilities of the Italian Center for Astronomical Archive (IA2) operated by INAF at the Astronomical Observatory of Trieste. This work exploits the Simbad, Vizier, and NASA-ADS
databases and the software TOPCAT \citep{topcat}.  This work has made use of data from the European Space Agency (ESA) mission Gaia (https://www.cosmos.esa.int/gaia), processed by
the Gaia Data Processing and Analysis Consortium (DPAC,
https://www.cosmos.esa.int/web/gaia/dpac/consortium). Funding
for the DPAC has been provided by national institutions, in particular the institutions participating in the Gaia Multilateral Agreement.
We acknowledge funding from MIUR Premiale 2016 MITiC. 
 \end{acknowledgements}

\begin{appendix}
\section{Data and results on individual stars}
\clearpage
\onecolumn
\begin{landscape}
\begin{longtable}{lccrccrrrr}
\caption{Information on the observed targets based on Gaia DR2.
}
\label{tab:A1}\\

\hline\hline
   Name        & Gaia ID&    Ra     & Dec 	  &  Gmag & BP-RP & Obs date & MJD-Obs& exp time &  \\
       &     & (J2000)    & (J2000)      &       &          &  &    &(s)&\\
\hline

ASCC 11  &  241730418805573760  &03:32:04.87& +44:57:45.4&  8.3675  &1.507  &2018-12-16  &58468.84  &1200  & \\
      & &   &    &    &  &2018-12-16    &58468.82   &1200    &  \\
\hline    
Alessi 1\_1   &402506369136008832	 & 00:53:15.44&+49:31:53.8& 9.824 &1.187   &2018-12-19  &58471.82 &1800  &   \\
         &  &   &    &    &  &2018-12-19    &58471.84   &1800    & \\
\hline     
Alessi 1\_2    &402505991178890752	 & 00:53:20.38& +49:28:49.7& 9.808& 1.176  &2018-12-19   &58471.87  &1800  &   \\
         & &   &    &    &  &2018-12-19    & 58471.89  &1800    &  \\
\hline
 Alessi 1\_3  &402867593065772288	& 00:54:51.06& +49:53:17.2&9.565 &1.186   & 2018-12-18 &58470.86  &1500  & \\
       &   &   &    &    &  &2018-12-18    & 58470.88  &1500    &  \\
\hline
 Alessi 1\_4  &402880684126058880 	 & 00:54:10.20& +49:40:08.9& 7.099& 1.349  &2018-12-19  & 58471.93 &1800  & \\
      &  &   &    &    &  &2018-12-19   &58471.95   &1800    &   \\
\hline
Alessi-Teusch 11 & 2184332753719499904 & 20:16:22.4& +52:06:18.4&7.099 &1.349 &2019-08-15    &58710.86  &690  &  \\
\hline
Basel 11b\_1 	 &3424056131485038592 & 05:58:08.10& +21:57:44.7&10.989 & 1.782  &2019-01-30   &58513.90  &3600  &   \\
\hline
Basel 11b\_2 	&3424055921028900736 & 05:58:10.12&+21:57:23.2&11.256 &1.876 & 2019-01-30 &58513.95  & 1800   \\
         & &   &    &    &  &2019-01-30     &58513.97   & 1800    \\
         & &   &    &    &  &2019-01-30    &58513.99   & 1800   &    \\
\hline
Basel 11b\_3  &3424057540234289408 	 & 05:58:18.16& +21:58:43.7&11.320 & 1.884 & 2018-12-20   &58472.19  & 1800  & \\
      &  &   &    &    &  &2018-12-20   &58472.21   & 1800   &   \\
         &    &   &    &    &  &2018-12-20     &58472.24   & 1800   &  \\
         &    &   &    &    &  &2018-12-20    &58472.26   & 1800   &  \\
\hline
COIN-Gaia 30  & 532533682228608384& 01:24:05.26& +70:25:25.1&8.352 & 1.762  & 2019-12-07  &58824.85  & 1400 & \\
\hline
Collinder 463\_1 &534207555539397888 & 01:33:49.45& +71:51:09.6&7.946 &1.632  & 2018-12-19  & 58471.80 &1200  & \\
\hline
Collinder 463\_2 & 534363067715447680 & 01:45:09.15& +71:53:25.3&7.725 & 1.572 &2018-12-17 &58469.81  &1200  & \\
\hline
Gulliver 18   & 1836389309820904064	& 20:11:43.9& 26:35:07.0&8.964 & 2.178  & 2019-08-10   &58705.88  &1380  &  \\
\hline
Gulliver 24	&430035249779499264 & 00:04:28.47& +62:42:04.4&10.494 & 1.927  & 2019-08-12   &58707.15  &1500  &\\
     &   &   &    &    &  &2019-08-12     &58707.16   &1500    & \\
\hline    
Gulliver 37	&2024469226291472000  & 19:28:18.44& +25:22:53.4&10.592  & 1.559  &2019-08-13  &58708.94 & 608 & \\
      & &   &    &    &  &2019-08-13   &58708.93   &1500    &\\
\hline     
NGC 2437\_1 & 3029609393042459392   & 07:41:36.9& -14:26:11.2&10.277 & 1.338  &2018-12-17  &58469.08  & 1800 &\\
        & &   &    &    &  &2018-12-17   &58469.10  &1800    &  \\
\hline     
NGC 2437\_2 &3029202711180744832   & 07:41:28.52& -14:54:17.5&10.293 &1.285   & 2018-12-17  &58469.12  &1800  &\\
         &  &   &    &    &  &2018-12-17     & 58469.15  &1800    &\\
\hline    
NGC 2437\_3   &3030364134752459904   &07:41:00.64& -14:12:08.4&10.417 &1.274  &2018-12-17  &58469.17&1800  & \\
          & &   &    &    &  &2018-12-17    & 58469.19  &1800    &\\
\hline    
NGC 2437\_4   &3029132686034894592 &07:42:47.85& -15:17:44.16&10.838 &1.302  &2018-12-18   & 58470.18&2400  &\\
            &  &   & &   & &2018-12-18       & 58470.20  &2400    &\\
\hline    
NGC 2437\_5  &3029156222454419072 &07:42:41.24& -14:59:51.4&10.948 &1.262  &2018-12-19    &58471.17  &2400  &\\
         &   &   &    &    &  &2018-12-19    &58471.20   & 2400   &  \\
\hline    
NGC 2437\_6  & 3029207006148017664 & 07:41:19.42& -14:48:47.5&9.877 & 1.332  &2018-12-18  &58470.10  &1500  &  \\
          &  &   &    &    &  &2018-12-18   &58470.12   &1500    &  \\
\hline    
NGC 2437\_7  & 3029226694277998080  &07:41:19.36& -14:40:59.7&9.975 &	1.322   &2018-12-08   &58470.14 &1500  &  \\
        &  &   &    &    &  &2018-12-08   & 58470.16  &1500    &   \\
\hline    
NGC 2509    &5714209934411718784    &08:00:44.36 &-19:06:59.4& 12.883&1.275 &2019-01-15    &58498.07  &2400  &  \\
          &  &   &    &    &  &2019-01-15    &58498.10   &2400    &  \\
        &  &   &    &    &  &2019-01-15    &58498.13   &2400    &  \\
 \hline   
NGC 2548\_1	&3064481400744808704 & 08:13:35.42& -05:53:02.04&9.377 & 1.098 &2018-12-17  &58469.24  &1200  &  \\
           &  &   &    &    &  &2018-12-17    & 58469.25  &1200    &  \\
\hline    
NGC 2548\_2	&3064537647636773760  & 08:12:37.24& -05:40:51.0&9.151 &1.111   &2018-12-18&58470.24  &1200  &\\
          & &   &    &    &  &2018-12-18    & 58470.25  &1200    & \\
\hline    
NGC 2548\_3   &3064579703955646976  	 &08:14:28.10& -05:42:16.09&9.187 &1.190 & 2018-12-19   &58471.23  &1200  &  \\
        &   &   &    &    &  &2018-12-19    & 58471.25  &1200    &  \\
\hline    
NGC 2548\_4   &3064486692144030336  	 & 08:13:40.44&-05:46:24.96&8.873 &0.786  & 2018-12-17  &58469.27  &1800  & \\
\hline    
NGC 7082\_1   & 1972288740859811072	  & 21:28:48.97&+47:06:54.2&8.171 &1.508   &2019-12-08  &58825.80 &1400  &  \\
\hline    
NGC 7082\_2 & 1972288637780285312	  & 21:28:34.58&+47:05:22.92&7.622 &1.470   &2019-12-07   &58824.81 &1400  &  \\
\hline
NGC 7209\_1 	&  1975004019170020736   & 22:05:09.94&+46:31:25.3&9.766 &1.323  & 2019-08-14 &58709.14  &1725  & \\
        &   &   &    &    &  &2019-08-14    &58709.16   &1725    &  \\
\hline     
NGC 7209\_2 	&1975002919658397568     & 22:05:17.63&+46:29:00.6&8.891 &1.734 & 2019-08-14  &58709.12  &2070  & \\
\hline    
Tombaugh 5\_1 	&473266779976916480   & 03:47:30.99&+59:02:50.8&11.163 & 2.123 &2018-12-20    &58472.11 &1800  &\\
           &  &   &    &    &  &2018-12-20     & 58472.09  &1800    &  \\
  &  &   &    &    &  &2018-12-20    &58472.15   &1800    & \\ 
  &  &   &    &    &  &2018-12-20     &58472.13   &1800    & \\
 \hline
 Tombaugh 5\_2 	&473275782228263296  & 03:48:32.98&+59:15:16.56&11.366 &2.147 &2019-01-15 &58497.06  &1800  &\\
   &  &   &    &    &  &2019-01-14   &58497.04   &1800    & \\ 
   &  &   &    &    &  &2019-01-15   &58498.00   &1800    & \\
\hline
Tombaugh 5\_3 	&473268424940932864 & 03:47:46.78&+59:05:36.6&11.138&2.028 &2019-01-14 &58498.03  &1800  &\\
   &  &   &    &    &  &2019-01-14   &58497.98   &1800    & \\ 
   &  &   &    &    &  &2019-01-14   &58497.02  &1800    & \\
           
\hline    
UPK 219   & 2209440823287736064 	& 23:30:29.72& +65:08:35.3&8.728 &1.734  &2019-12-08 &58825.84  &1400  & \\
\hline    
\multicolumn{9}{c}{Comparison clusters} \\
Collinder 350\_1 &4372743213795720704	& 17:46:24.88& +01:02:39.7&5.957 &1.869   &2018-08-20   &58350.87 &300 &  \\
           & &   &    &    &  &2018-08-20     &58350.88   &300    & \\
\hline    
Collinder 350\_2 &  4372572888274176768	& 17:48:43.82& +01:09:51.1&8.421 & 1.485& 2018-08-20  &58350.88  &1380  & \\
\hline     
NGC 2682\_1 &604921512005266048 	 & 08:51:26.17&+11:53:51.9&10.202 &1.238  &2020-02-02 &58881.07  &1500  & \\
          &   &   &    &    &  &2020-02-02     & 58881.08  &1500    &\\
\hline    
NGC 2682\_2 &604920202039656064 	 & 08:51:59.51&+11:55:04.8 &10.205 & 1.243 &2020-02-02   &58881.10  &1500  &\\
           &  &   &    &    &  &2020-02-02    &58881.12   &1500    &\\
\hline    
NGC 2682\_3  & 604904950611554432	 & 08:51:50.19&+11:46:06.9&10.511 &1.110&2020-02-02    &58881.14 &1500  &\\
   &  &   &    &     &  &2020-02-02    &58881.16  &1500 &\\
    &   &   &    &    &  &2019-12-08    &58825.19   &1500    &  \\ 
    &   &   &    &    &  &2019-12-08     &58825.21   &1500    &   \\ 
    &   &   &    &    &  &2019-12-08     &58825.22   &1500    &  \\ 
    &   &   &    &    &  &2020-03-11    &58919.04   &1500    &  \\ 
    &   &   &    &    &  &2020-03-11   &58919.06  &1500    &   \\
\hline              
NGC 2682\_4  &604917728138508160	 & 08:51:23.76&+11:49:49.3&11.231 &1.073  &2020-03-10    &58918.99  &1420  & \\
     &         &   &    &    &  &2020-03-11   &58919.00   &1420    & \\
   & &   &    &    &  &2020-03-11     &58919.02   &1420    & \\
\hline
\end{longtable}

\end{landscape}

\clearpage
\twocolumn

\clearpage
\onecolumn
\begin{landscape}
\begin{longtable}{lccccccccccccccccc}
\caption{Final atmospheric parameters of observed targets}
\label{tab:A2}\\
\hline\hline
  Name        &     Gaia ID & $T_{\textrm{eff}}$ &  $\sigma$  $T_{\textrm{eff}}$&   $\log$ g &	 $\sigma$   &  [Fe/H]& $\sigma_1$& $\sigma_2$  & $\log \epsilon$  &  std &  Nlines &  $\log \epsilon$  & std & Nlines&    v$_{micro}$& $\sigma$     \\
             &    & (K)  &  & (dex)  &$\log$ g  & (dex)  & [Fe/H]      &[Fe/H]  &Fe I & &Fe I &Fe II &  &Fe II &(km~s$^{-1}$)&v$_{micro}$ \\
\hline
ASCC 11  &241730418805573760 & 5250 &70&2.15&0.10 &-0.14&0.05 &0.06 &7.360& 0.159   & 46 &7.360 & 0.210& 10&2.4&0.3 \\
Alessi 1\_1 &402506369136008832	 &5000&70 &2.65&0.15 &-0.10&0.05&0.05 &7.392 &0.181&74 &7.391 & 0.136&16& 1.5& 0.1 \\
Alessi 1\_2 &402505991178890752	 &5200&70 &3.20&0.10 &0.08&0.10&0.10 &7.583 &0.135&75 &7.587& 0.128&15& 1.5&0.2 \\
Alessi 1\_3 & 	402867593065772288  &5250&65 &3.27&0.10 &0.07&0.06 &0.06 &7.576 &0.105&75 &7.575& 0.098&15& 1.6&0.1 \\ 
Alessi 1\_4 & 402880684126058880	 &5120&60 &3.09&0.06 &-0.05&0.10&0.10 &7.470 &0.122&72 &7.476& 0.089&16& 1.6&0.1 \\
Alessi-Teusch 11 &2184332753719499904  &4560&50 &2.10&0.15 &-0.19&0.05&0.05&7.310&0.172&75 &7.311& 0.219&16& 2.1&0.2 \\
Basel 11b\_1 &3424056131485038592	 &5180&50 &3.15&0.13 &0.00&0.05&0.05 &7.500 &0.093&68 &7.507& 0.127&13& 2.2&0.1 \\
Basel 11b\_2 &3424055921028900736&5220&30 &3.07&0.10&0.02&0.03&0.03&7.525 &0.141&72 &7.529& 0.135&15& 2.2&0.2 \\
Basel 11b\_3  &3424057540234289408 	 &4950&50&2.83& 0.10 &-0.04&0.05 &0.05  &7.464&0.136&77 &7.467& 0.116&15& 2.1&0.1 \\
COIN-Gaia 30 &532533682228608384	&5200&50 &3.40&0.05 &0.03&0.05&0.05 &7.538 &0.158 &71&7.552& 0.163&15& 2.3&0.2 \\
Collinder 463\_1 &534207555539397888 &4730&30 &2.12&0.10 &-0.20&0.05&0.05 &7.310 &0.130&72 &7.310& 0.180&16& 2.5&0.1 \\
Collinder 463\_2 & 534363067715447680 &4730&30 &2.30&0.15 &-0.10&0.05&0.05 &7.410 &0.131&70 &7.410& 0.207&16& 2.4&0.1 \\
Gulliver 18  & 1836389309820904064	&4590&100 &2.60&0.17 &-0.10&0.04&0.05 &7.409 &0.212 &76&7.395& 0.281&16& 2.8&0.3 \\
Gulliver 24 &430035249779499264 	&4450&50 &2.50&0.15 &-0.18&0.03&0.03 &7.317&0.145&74 &7.316& 0.240&16& 2.0&0.2 \\
Gulliver 37  & 2024469226291472000	&4850&50 &3.65&0.12 &0.10&0.03&0.04 &7.602 &0.213&62 &7.604& 0.403&10& 0.8&0.6 \\
NGC 2437\_1 &3029609393042459392  &5050&50 &2.77&0.12 &0.04&0.05&0.05 &7.546 &0.142&64&7.548& 0.109&12& 2.1&0.1 \\
NGC 2437\_2 &3029202711180744832  &5250&75 &3.32&0.10 &0.07&0.05&0.05 &7.572&0.140&62 &7.578& 0.171&13& 2.2&0.2 \\
NGC 2437\_3 & 3030364134752459904  &5300&110 &3.13&0.10 &-0.05&0.05&0.05 &7.450 &0.176&62 &7.447& 0.149&13& 2.3&0.2 \\
NGC 2437\_4   & 3029132686034894592 &5085&65 &2.90&0.10 &0.00&0.05&0.05 &7.508 &0.111&62 &7.498& 0.089&15& 1.7&0.1 \\ 
NGC 2437\_5 &3029156222454419072  &5030&75 &2.85&0.15 &0.00&0.10&0.10 &7.503 &0.163&74 &7.505& 0.161&16& 1.7&0.1 \\
NGC 2437\_6 &3029207006148017664   &4990&65 &2.72&0.10 &0.00&0.05&0.06 &7.443 &0.238&72 &7.444& 0.266&15& 2.1&0.1 \\
NGC 2437\_7 & 3029226694277998080   &4650&90 &2.27&0.20 &-0.07&0.08&0.08 &7.440&0.216&73 &7.440& 0.272&16& 1.2&0.1 \\
NGC 2509 & 5714209934411718784  &4705&40 &2.53&0.20 &-0.10&0.05&0.06 &7.394 &0.213&72 &7.392& 0.330&16& 1.5&0.3 \\
NGC 2548\_1	& 3064481400744808704  &5370&70 &3.67&0.15 &0.00&0.05&0.05 &7.507 &0.131&60 &7.504& 0.155&15& 2.0&0.2 \\
NGC 2548\_2	&3064537647636773760 &5050&50 &2.65&0.10 &-0.02&0.04&0.04 &7.479 &0.068&62 &7.480& 0.101&15& 1.6&0.1 \\
NGC 2548\_3    &3064579703955646976	 &4930&50 &2.70&0.10 &-0.01&0.05&0.05 &7.490 &0.079 &61&7.491& 0.083&13& 1.6&0.1 \\
NGC 2548\_4 	&3064486692144030336  &5200&50 &3.18&0.10 &-0.07&0.07&0.07 &7.434 &0.110&62 &7.439& 0.142&14& 0.4&0.1 \\
NGC 7082 	&1972288740859811072  &4994& 50 &2.25& 0.1 &-0.15&0.05  &0.07  &7.360 &0.144&63 &7.361& 0.199&12& 3.0&0.2   \\
NGC 7209\_1 	&1975004019170020736    &4880&50 &2.35&0.15 &0.00&0.03&0.07 &7.439 &0.110 &74&7.433& 0.155&16& 1.7&0.2 \\
NGC 7209\_2 	 &1975002919658397568 &4600&30 &2.79&0.15 &-0.07&0.05&0.05 &7.433 &0.136 &67&7.435& 0.316&16& 2.5&0.2 \\
Tombaugh 5\_1 	 & 473266779976916480 &5010&50 &3.17&0.15 &0.04&0.05&0.05 &7.543 &0.166 &72&7.546& 0.185&15& 2.2&0.1 \\
Tombaugh 5\_2 	 & 473275782228263296 &4900&50 &2.31& 0.1 &-0.07&0.05  &0.01   &7.438 &0.080 &74&7.436& 0.073&14& 2.0&0.2 \\
Tombaugh 5\_3 	 & 473266779976916480 &5150&50 &3.08&0.15 &0.07&0.05 &0.05 &7.570 &0.095 &69&7.576& 0.147&15& 2.2&0.2 \\
UPK 219    & 2209440823287736064 	&5203&150 &3.01&0.10 &0.02&0.05&0.05 &7.528 &0.166&74 &7.530& 0.283&16& 2.7&0.2 \\
\hline    
Collinder 350\_1   &4372743213795720704	&4200&50 &1.30&0.20 &-0.40&0.08&0.08 &7.120 &0.149&71 &7.130& 0.256&16& 2.2&0.1 \\
Collinder 350\_2  &4372572888274176768 	&5300&50 &3.15&0.1 &0.02&0.05 &0.05  &7.521 &0.079 &71&7.526& 0.095&15& 1.6&0.1 \\
NGC 2682\_1    &604921512005266048 	 &4687&50 &2.37&0.07 &-0.05&0.05&0.05 &7.454 &0.174 &75&7.453& 0.181&16& 1.5&0.2 \\
NGC 2682\_2    &604920202039656064 	 &4900& 110 &2.76& 0.1 &0.02&0.05 & 0.05  &7.520 &0.132 &69&7.528& 0.168&15& 1.7& 0.2 \\
NGC 2682\_3  	&604904950611554432 &5000&50 &2.77&0.1 &-0.03&0.03  & 0.03  &7.467 &0.096&67 &7.467& 0.082&14& 1.2& 0.2 \\
NGC 2682\_4  	&604917728138508160 &5195&50 &3.25&0.10 &0.00&0.05 &0.03  &7.501 &0.108&67 &7.502& 0.164&14& 1.3&0.2 \\
SUN  	        &~~~ &5770&40 &4.44&0.08 &-0.03&0.03&0.03 &7.477 &0.077&76 &7.477& 0.083&16& 1.0&0.1 \\    
\hline

\end{longtable}
\tablefoot{$\sigma_1 [Fe/H]$ is the error from metallicity measurement, $\sigma_2 [Fe/H]$ is the deviation of iron abundance combined to the uncertainties of EW and error in metallicity determination }
\end{landscape}

\clearpage
\twocolumn

\end{appendix}

\end{document}